\documentclass[aps,pre,twocolumn,superscriptaddress,floatfix]{revtex4}
\usepackage{graphicx}
\usepackage{times}
\usepackage[dvips,unicode,colorlinks,linkcolor=blue,citecolor=blue,urlcolor=blue]{hyperref}
\usepackage{amssymb}
\begin{document}

\title{Two-phase stretching of molecular chains}

\author{Alexander V. Savin}
\email[]{asavin@center.chph.ras.ru}
\affiliation{Semenov Institute of Chemical Physics, Russian Academy of Sciences, Moscow 119991, Russia}

\author{Irina P. Kikot} 
\affiliation{Semenov Institute of Chemical Physics, Russian Academy of Sciences, Moscow 119991, Russia}

\author{Mikhail A. Mazo}
\affiliation{Semenov Institute of Chemical Physics, Russian Academy of Sciences, Moscow 119991, Russia}

\author{Alexey V. Onufriev}
\affiliation{Departments of Computer Science and Physics, 2160C Torgersen Hall,
             Virginia Tech, Blacksburg, VA 24061, USA}

\begin{abstract}
While stretching of most polymer chains 
leads to rather featureless force-extension 
diagrams, some, notably DNA, exhibit non-trivial behavior 
with a distinct plateau region. Here we propose a unified theory that 
connects force-extension characteristics of the polymer chain with the 
convexity properties of the extension energy profile of its individual 
monomer subunits. 
Namely, if the effective  monomer deformation energy as a function 
of its extension has a non-convex (concave up) region, 
the stretched polymer chain separates into two phases: the weakly 
and strongly stretched monomers.  Simplified planar and 3D polymer 
models are used to illustrate the basic principles of the proposed model. 
Specifically, we show rigorously that when the 
secondary structure of a polymer is mostly due to weak non-covalent 
interactions, the stretching is two-phase, and the force-stretching diagram 
has the characteristic plateau. We then use realistic coarse-grained models to 
confirm the main findings and make direct connection to the microscopic structure of the monomers. 
We demostrate in detail how the two-phase 
scenario is realized in the $\alpha$-helix,
and in DNA double helix. The predicted plateau parameters  
are consistent with single molecules experiments.
Detailed analysis of DNA stretching 
demonstrates that breaking of Watson-Crick bonds is not necessary for the 
existence of the plateau, although some of the bonds 
do break as the double-helix extends at room temperature. 
The main strengths of the 
proposed theory are its 
generality and direct microscopic connection. 
\end{abstract}

\maketitle

\section{Introduction\label{s1}}

When pulled by the ends, a flexible linear polymer
first undergoes entropic elongation
where the work done by the stretching force reduces the conformational entropy
of the chain \cite{Khokhlov1994,Marko95}. 
In this well-understood \cite{Marko95} weak 
extension regime the polymer obeys Hooke's law and its
elastic properties are ``universal", in that they are
insensitive to details of the chemical
structure and interactions within its monomers.
As the polymer chain is extended further, and its
end-to-end distance becomes comparable to the chain contour length,
the intrinsic elasticity due to deformation and interaction of
individual monomers begins to dominate the extension response \cite{Rief97}.
Since short-scale chemical structures of real polymers differ
substantially, as do their observed responses to strong tension forces, 
one wonders if polymer stretching in this regime 
can still be described by a universal principle? The question is important.  
Biopolymers such as DNA are subjected to a range of 
mechanical manipulations within the cell, they  
may change their conformations and
undergo unexpected structural transitions \cite{Smith,Cluzel,Allemand98}.
Knowledge of elastic properties of biopolymers
is required to understand the structural 
dynamics of many important cellular processes \cite{Garcia07,Bustamante04,Nelson99}. These properties can now be measured quite accurately by  
modern experimental techniques such as atomic force microscopy
and optical tweezers. 

For DNA \cite{Cluzel, Smith,Williams2004, McCauley2008,Gross2011} and 
polypeptides \cite{Schwaiger2002, Afrin2009, Ritort2006} 
these experiments have revealed several peculiar features. 
When extended, the (double-stranded) DNA molecule exhibits the following 
behavior: until the end-to-end distance reaches 0.9 of the contour length,  
the stretching process is  well described by established 
phenomenological models \cite{Bustamante,Smith,Marko95}. But then, 
when the molecule is subjected to forces of $65\div120$ pN (depending of experimental conditions), 
a sudden structural transition occurs, in which the chain stretches
up to 70\%  beyond its canonical B-form contour length. 
The extension force remains almost constant in this regime, 
which is manifested by a characteristic plateau on the experimental 
force-extension curve.  
Similar single molecule stretching experiments have also been performed 
on polypeptide molecules \cite{Schwaiger2002, Afrin2009, Ritort2006}.
It was found that simple \footnote{Some common polypeptides have rather complex secondary structures, and many
features of their force-extension diagrams stem from gradual
disruption of the secondary structure elements.}  
helical polypeptide structures such as synthetic alpha-helices \cite{Afrin2009} and myosin molecules \cite{Schwaiger2002},  exhibit 
a force-extension plateau similar to that seen in DNA 
stretching experiments. In contrast, these features are not 
observed in many "non-biological" polymers such as polyethylene.    

Various microscopic models were proposed to explain these 
observations on a case-by-case basis. 
For example, 
force-extension plateau observed in single DNA molecule experiments (sometimes called the over-stretching plateau) is often 
explained by 
gradual un-zipping (force-induced melting) 
of the double helix in which Watson-Crick (WC) hydrogen 
bonds between base-pairs break
 \cite{Rouzina2001_1, Rouzina2001_2, Williams2004, McCauley2008,vanMameren2009, Gross2011},
An alternative explanation involves 
cooperative transition of the whole structure into a new form called S-form 
where
WC bonds remain intact \cite{Cluzel, Lebrun1996}, but the helix unwinds to form a straight ladder. 
In the case of polypeptides, force-extension plateau is attributed to 
alpha-helix unwinding \cite{Rief98, Schwaiger2002, Zegarra2009}. 
Phenomenological descriptions based on various assumptions about 
stable monomer states were also proposed \cite{Storm03}.  
Still, no universal, microscopically based mechanism exists that can 
explain why some polymers 
do and some 
do not exhibit a plateau in force-extension experiments. Here we propose 
such a mechanism and show how the stretching properties of the polymer depend on the balance between valent and non-valent interactions on the 
scale of individual monomers. 

\section{Results \& Discussion\label{s2}}

\subsection{The general mechanism of polymer stretching under tension}
 Consider a linear polymer chain of $N \gg 1$ identical interacting sites (monomeric units). The {\it effective} site deformation energy 
$E(\Delta l)$ can be defined as follows. Consider a configuration of the 
chain in which each site is stretched by the same amount $\Delta l$. Then 
$E(\Delta l)$ is simply 
the total deformation energy of the chain divided by $N$. 
Here we show that the 
shape of the effective site deformation energy $E(\Delta l)$ 
determines force-induced stretching behavior 
of the chain in the general experimental scenario when the force 
is applied to the chain's ends and no restrictions are imposed 
on deformations of individual sites.   
\begin{figure}[bt]
\includegraphics[angle=0, width=0.9\linewidth]{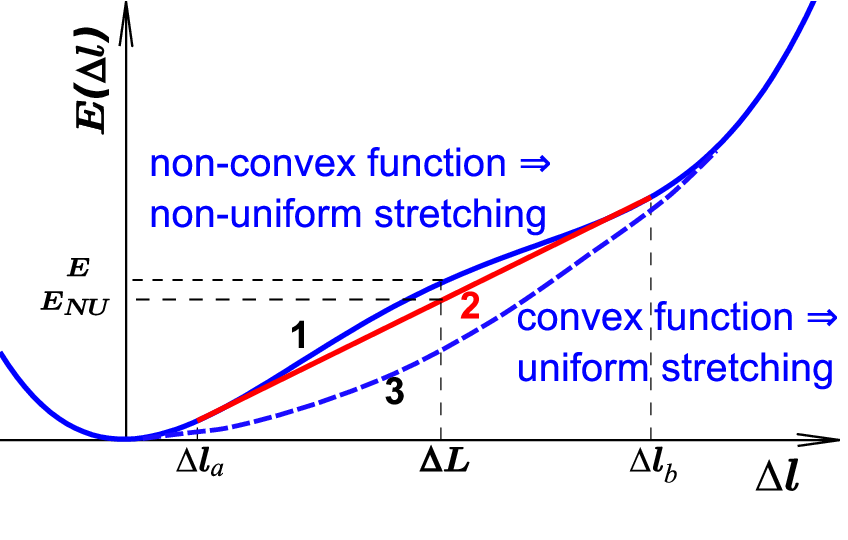}
\caption{\label{fig1}
Two distinct forms of the effective 
site deformation energy $E(\Delta l)$ 
of an individual monomeric unit (site) that lead to qualitatively 
different stretching scenarios of the linear polymer chain of $N$ sites. 
The site extension is $\Delta l$. 
Curve 1: A non-convex energy function (between $\Delta l_a$ and $\Delta l_b$)  
leads to non-uniform, two-phase  
stretching of the chain. Some sites extend weakly by $\Delta l_a$, and 
some strongly by $\Delta l_b$.
The chain extension follows 
the convex hull (red line 2) of $E(\Delta l)$, with only the relative 
fraction of weakly extended sites changing as the chain extends. 
The average (per site) 
energy of non-uniform extension $E_{NU}(\Delta L)$ is 
less than that of the corresponding uniform $E(\Delta L)$ extension. 
Curve 3: A convex function $E(\Delta l)$ 
leads to uniform extension of all the sites. 
} 
\end{figure}

In what follows we will use the following convention. 
Extension of a single monomeric site or, equivalently, 
that of each site of a uniformly stretched chain is denoted by $\Delta l$. 
In general, including the case of non-uniform deformation, 
extension of site $i$ is denoted by 
$\Delta l_i$. We use $\Delta L$ for the mean per site deformation 
$\Delta L = N^{-1}\sum_{i=1}^N \Delta l_i$.  
If the function $E(\Delta l)$ is convex down, 
line 3 in Fig.~\ref{fig1}, 
the most favorable 
structure of the chain with fixed total deformation $\sum_i \Delta l_i$ 
corresponds to each site $i$ stretched by the same amount $\Delta l_i \equiv \Delta l$, 
see appendix \ref{sa} for details. 
The simplest example of such a polymer model is a chain of beads connected 
by harmonic springs.  
The total deformation energy of the chain in the case of convex $E(\Delta l)$  is $NE(\Delta l)$. 
Non-uniform stretching is energetically unfavorable in this scenario because 
any putative decrease in the total energy from under-stretching 
of a group of sites   
would be offset by a larger increase in the energy of 
the remaining sites that would 
have to over-stretch to keep the total deformation 
of the chain constant. 
        In contrast, if the effective site extension energy 
is non-convex over some interval, 
(curve 1, Fig.~\ref{fig1}), two-phase stretching becomes more favorable 
energetically: one part of chain consists of 
$pN$ sites ($0 < p < 1$)  stretched "strongly" by $\Delta l_b$, and another part consists of the rest $(1-p)N$ sites  stretched "weakly" by $\Delta l_a < \Delta l_b$.  
Qualitatively, this is because the decrease of the chain energy
(relative to the uniform stretching scenario) resulting 
from under-stretching of a 
group of sites is larger than the gain from over-stretching of the remaining
sites. A detailed quantitative analysis is presented in the appendix \ref{sa}.   
Briefly, the mean deformation per site 
in  this case is $\Delta L=p\Delta l_b + (1-p)\Delta l_a$, and 
the total energy of the chain in this non-uniform (NU) case equals 
$NE_{NU}(\Delta L) = N\left( (1-p)E(\Delta l_a)+pE(\Delta l_b) \right)$ (contribution from phase boundary can be neglected for long chains, 
$N \gg 1$, typically used in experiment \cite{vanMameren2009} ).  Since the  
function $E(\Delta l)$ is non-convex, $NE_{NU}(\Delta L)$ is less than  
$NE(\Delta L) = NE((1-p)\Delta l_a+p\Delta l_b)$ -- 
the total energy of the chain in the uniform case with 
the same total deformation, Fig.~\ref{fig1}. 
In the non-uniform regime, the chain extension is achieved via
change in the relative fraction $p$ of the strongly stretched sites,
not by extension of individual sites. 
As $p$ increases from 0 to 1,
the mean deformation $\Delta L=(1-p)\Delta l_a + p \Delta l_b$ 
depends on $p$ linearly and ranges from $\Delta l_a$ to $\Delta l_b$. 
The average per site chain energy $E_{NU}(\Delta L)$ also depends on 
$p$ linearly, and ranges form $E(\Delta l_a)$ to $E(\Delta l_b)$ - that is the stretching process is described by a straight line connecting 
points $\Delta l_a$ and $\Delta l_b$ - see Fig.~\ref{fig1}, red line.
The tension force ${dE_{NU}}/{d\Delta L }$ thus remains constant,
and the characteristic plateau in the force-extension diagram appears.

Real polymer chains may appear more complex,  
but the chain structure is always stabilized by interactions of two types: 
"strong" valent and ``weak" non-valent.
The former describes bond, angle and torsion deformations. The latter 
corresponds to ``soft" non-valent interactions. 
These interactions include various combinations of electrostatic 
and van der Waals interactions, hydrogen bonds in polypeptide alpha-helix and stacking
interactions between neighboring base pairs in DNA. The main feature of 
realistic non-valent interactions potentials $W(r)$ is the existence of  
inflection points. 
If non-valent interactions contribute significantly 
to the extension energy $E(\Delta l)$, the function $E(\Delta l)$ may also have an inflection point and hence a non-convex region as in Fig.~\ref{fig1}, 
leading to a plateau in the force-extension diagram.

We thus propose a general mechanism for the observed 
two-phase stretching of liner polymers based on 
convexity properties of the effective potential energy of the 
monomeric units of the polymer. As 
we will demonstrate, the mechanism is able to explain  
the existence of force-extension plateaux for very different 
types of polymers.
\begin{figure}[tb]
\includegraphics[angle=0, width=1.\linewidth]{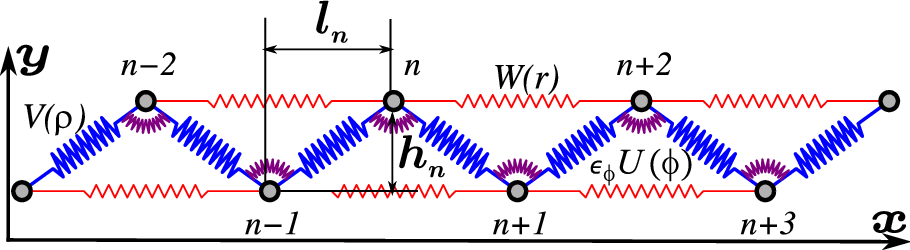}
\caption{\label{fig2}
Schematic of a 2D zigzag polymer chain. $\epsilon_\phi U(\phi)$ and $V(\rho)$ are 
the valent angle bending and bond stretching potentials; $W(r)$ is the 
non-valent interaction between next-nearest neighbors. The longitudinal step  
of site $n$ is $l_n$, the transverse step is $h_n$.}
\end{figure}
 
\subsection{Stretching of a 2D zigzag molecular chain}

We begin by exemplifying the proposed mechanism of non-homogeneous 
two-phase stretching on a 2D zigzag chain, Fig.~\ref{fig2}.  While arguably 
among the simplest polymer geometries, it is often found in 
real polymers: for example, polyethylene (PE) molecule has a stable plane 
conformation of {\it trans-zigzag}.
The 2D zigzag form is also common in  hydrogen-bonded chains
$\cdots$X--H$\cdots$X--H$\cdots$X--H$\cdots$ of halides, where X=F, Cl, Br, I.
\begin{figure*}[tb]
\begin{center}
\includegraphics[width=.8\linewidth]{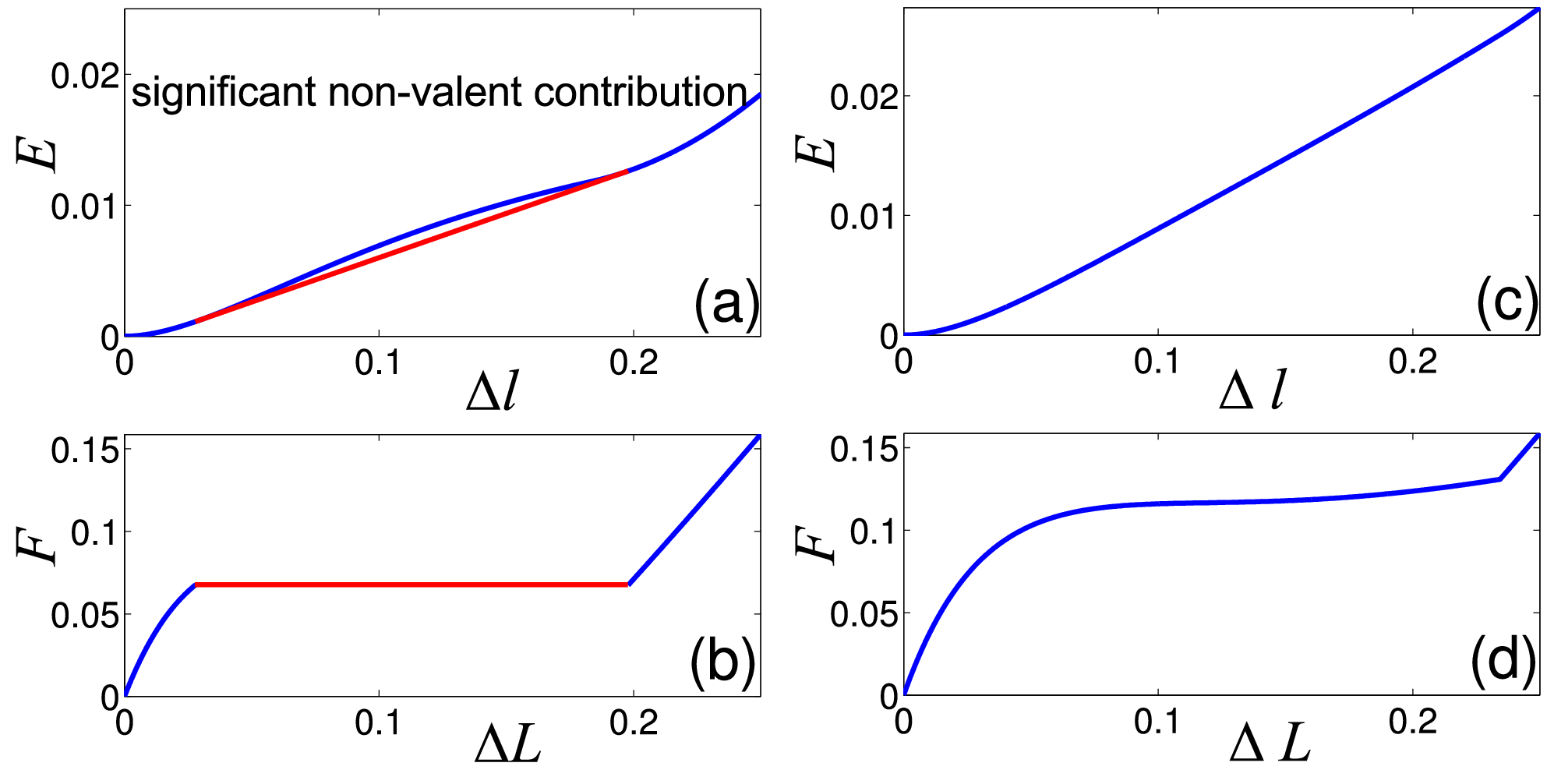}
\end{center}
\caption{\label{fig3}
(a, c):  
2D zigzag effective site deformation energy $E(\Delta l)$ as a function of site extension $\Delta l$ from equilibrium.   
Convex hull of $E(\Delta l)$ (red line in a) represents
two-phase stretching energy per one site, $E_{NU}(\Delta L)$. (b,d): 
Dependence of the tension force on mean site extension $\Delta L$.  
Angle deformation stiffness $\epsilon_\phi$, Fig. \ref{fig2}, is varied. 
Left: $\epsilon_\phi=0$.  Right: $\epsilon_\phi=0.02$. 
}
\end{figure*}

We consider a dimensionless model of 2D zigzag chain, 
see  section \ref{s4} and appendix \ref{sb} for details. 
In the limiting case of zero angle bending potential $\epsilon_{\phi} =  0$, 
Fig.~\ref{fig2}, only 
non-valent interactions between next nearest neighbors determine 
elastic response of the chain.
Numerical simulations of the zigzag with $N \gg 1$  
units, see section \ref{s4}, show that in this case the chain 
stretching is accompanied by monotonous increase of the bond angle;  
once the mean extension reaches $\Delta L=0.18$, the zigzag straightens 
out completely into a line (the zigzag angle is  $\phi=180^\circ$), then valent bonds begin to stretch.
The corresponding effective site deformation energy $E(\Delta l)$
of a single monomer site is shown in Fig.~\ref{fig3} (a). 
The function $E(\Delta l)$ 
is not convex; its 
convex hull is given by a tangent line at points $\Delta l_a=0.04$ and 
$\Delta l_b=0.21$. According to our general mechanism, a fraction of 
the zigzag sites is expected to be  
in  the weakly extended state with the longitudinal step
$l_n = l_0+\Delta l_a$, while the rest will be in the strongly extended 
state with the step $l_n = l_0+\Delta l_b$.
The tension force $F=dE_{NU}/d\Delta L$ remains constant between $\Delta l_a$ and 
$\Delta l_b$, 
and the force-extension dependence
$F(\Delta L)$ has the typical plateau,  Fig.~\ref{fig3} (b). 
        The zigzag effective 
site extension energy $E(\Delta l)$ remains non-convex until 
the strength of the angle potential reaches a critical value 
($\epsilon_\phi=0.015$). At this point, energetic 
benefit from non-uniform stretching relative to uniform stretching 
vanishes. As the angle bending potential becomes even stiffer, 
its relative contribution to chain stretching overwhelms that 
of the weak non-valent interactions that give rise 
to the non-convex behavior seen in Fig.~\ref{fig3} (a). The 
effective site extension energy function becomes convex, Fig.~\ref{fig3} (c), 
and the stretching behavior of the polymer is essentially that of a harmonic 
spring -- single phase, uniform.  These scenarios are 
further illustrated in the Appendixes for a zigzag chain of $N=400$ ``atoms".

 Thus, two-phase stretching and force-extension plateaux 
can be expected to  occur in molecular chains
in which secondary structure is supported by weak non-valent interactions.
On the other hand, if the secondary structure is due mainly 
to angle  deformation, then the stretching will be uniform. 
Such a scenario is typical for polyethylene (PE) trans-zigzag \cite{Manevitch89}.

\subsection{Stretching of the $\alpha$-helix}

\begin{figure}[tb]
\includegraphics[angle=0, width=1\linewidth]{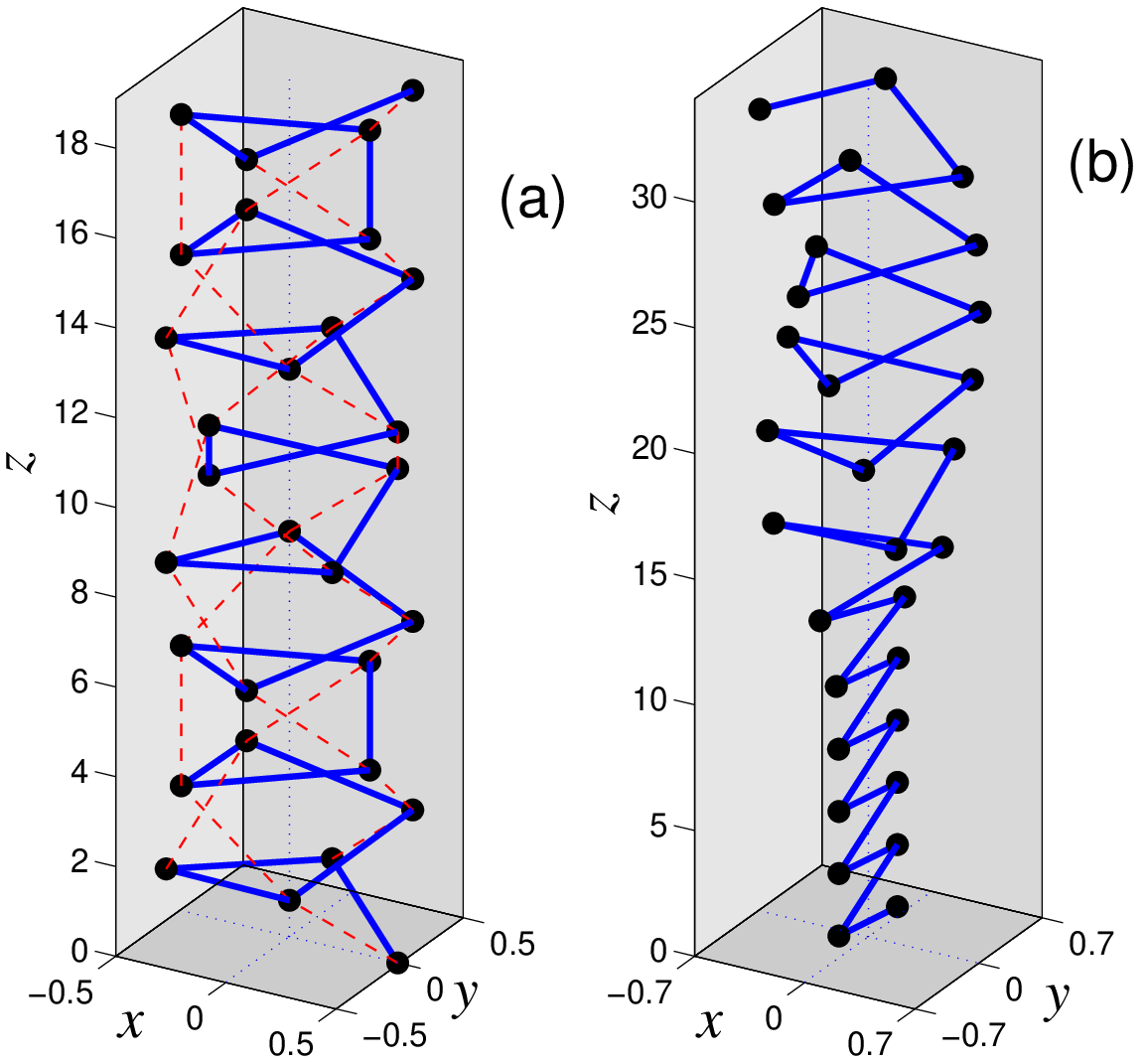}
\caption{\label{fig7}
(a) Schematic  of an $\alpha$-helix. Dimensionless units.
Helix monomeric sites are shown in their equilibrium positions,
solid blue lines denote rigid valent bonds while red dotted lines designate 
soft hydrogen bonds.
(b) Transition between strongly ( bottom half ) and weakly (top half ) 
stretched parts of the helix. 
        }
\end{figure}

Consider a 3D molecular chain 
corresponding to an ideal \cite{Khokhlov1994} 
$\alpha$-helix, Fig.~\ref{fig7} (a). 


\begin{figure}[tbh]
\includegraphics[width=1\linewidth]{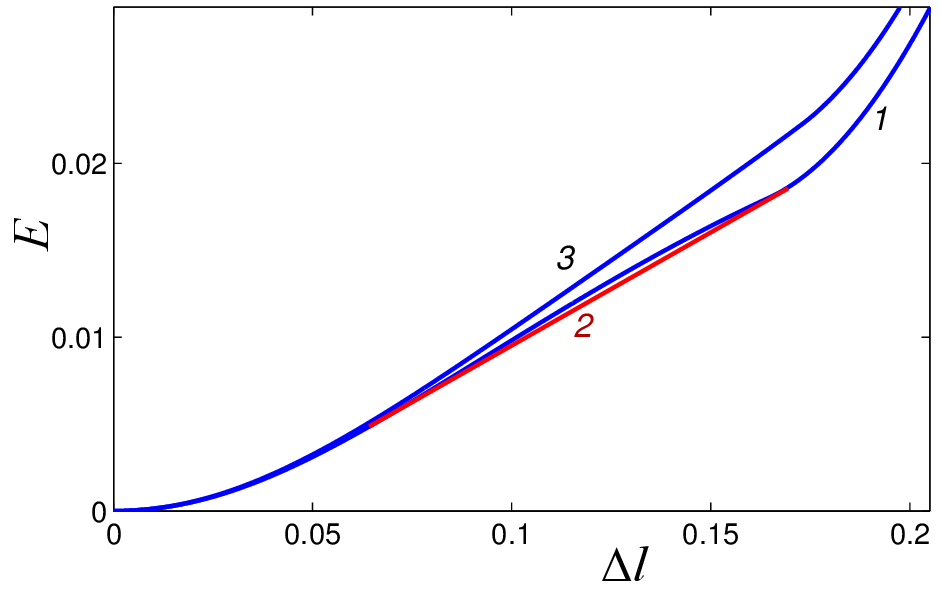}
\caption{\label{fig8}
Effective site extension energy $E$ 
of the helix 
as a function of longitudinal site extension $\Delta l$ from equilibrium.
Torsion rigidity is $\epsilon_\theta=0$ (curve 1), $\epsilon_\theta=0.002$ (curve 3). The red line 2 is the convex hull of curve 1.
        }
\end{figure}
Here, the softest valent potential is the torsional potential; 
we vary its 
relative contribution $\epsilon_\theta$ to the total energy  
while keeping the other parameters fixed, see section \ref{s4} and appendix \ref{sc}.
 Without the torsional rigidity ($\epsilon_\theta=0$),
the helix is stabilized only by hydrogen bonds, connecting site $i$ with sites 
$(i+3)$  and $(i-3)$ \cite{Khokhlov1994,Christiansen97,Savin2000}.
The effective site energy  $E(\Delta l)$ 
is shown in Fig.~\ref{fig8}. Upon stretching, 
the helix's angular step (deformation) monotonously increases 
and reaches its maximum value $180^\circ$ when $\Delta l=0.16$ (plane zigzag).
The function $E(\Delta l)$ is not convex, Fig.~\ref{fig8} (a). 
Its convex hull is described by a tangent at 
points $\Delta l_a=0.06$, $\Delta l_b=0.17$. According to our 
general mechanism, a fraction of the helix 
is in the weakly extended state with the longitudinal step
$l_0+\Delta l_a$, while the rest is in the strongly extended (plane zigzag) state with the longitudinal step $l_0+\Delta l_b$.
As long as $\Delta l_a<\Delta L<\Delta l_b$, 
the tension in the helix $F=dE_{NU}/d\Delta L$ remains constant, leading to the 
characteristic plateau in the force-extension diagram.
If the torsional rigidity is increased, the non-convex shape of 
$E(\Delta l)$ is preserved until $\epsilon_\theta=0.0015$ is reached. Once
$\epsilon_\theta>0.0015$, the function $E(\Delta l)$ becomes convex -- see Fig.~\ref{fig8}. In this regime, only uniform stretching of the helix is 
possible.  
\begin{figure}[tb]
\includegraphics[width=1\linewidth]{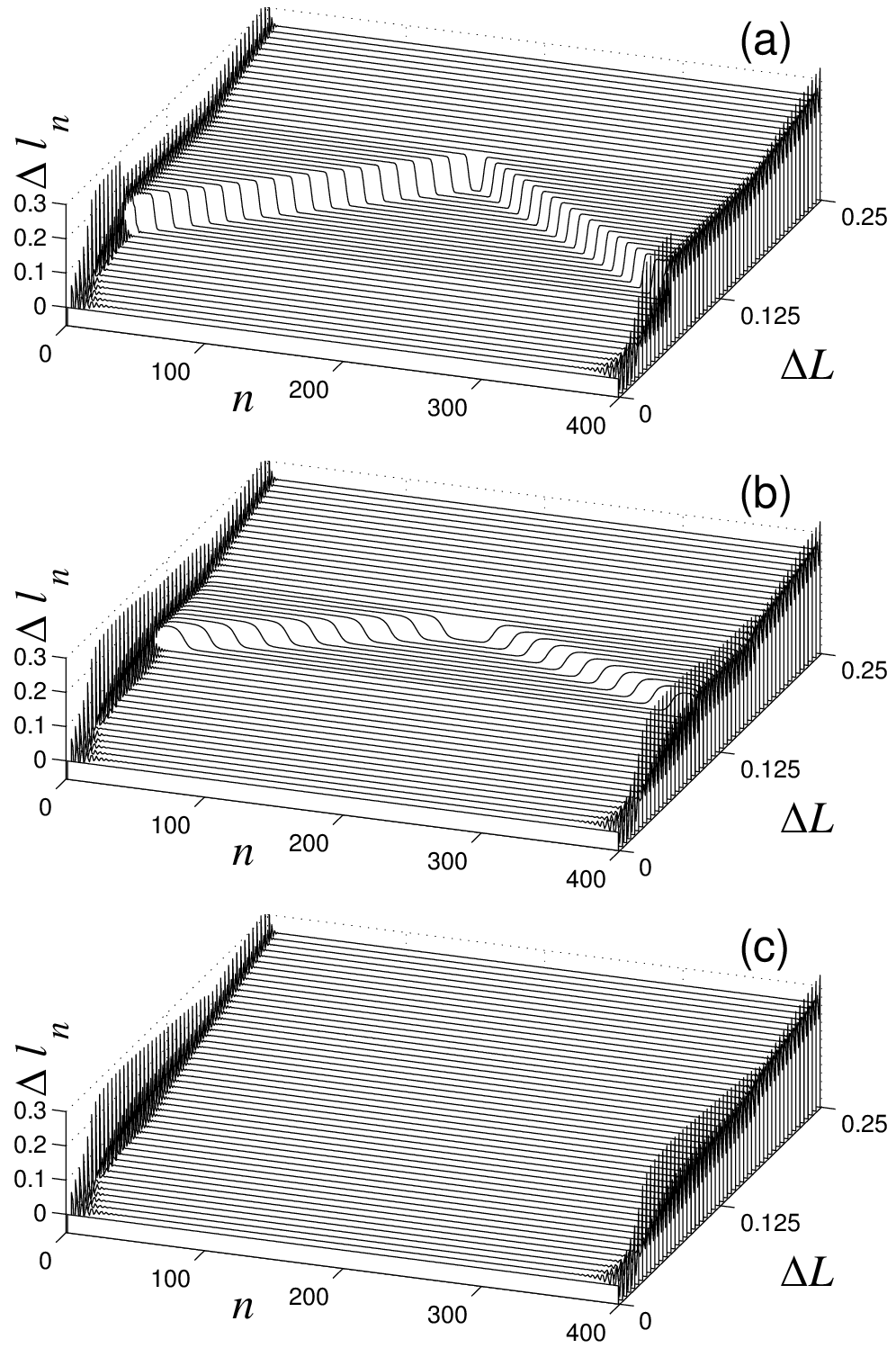}
\caption{\label{fig10}
Distribution of longitudinal extensions  $\Delta l_n$ of individual sites
in a stretched helix of $N=400$ sites as a function of the mean (per site ) 
helix extension $\Delta L = N^{-1}\sum_n^N \Delta l_n$. Three values of the 
torsional rigidity are considered: 
(a) $\epsilon_\theta=0$, (b) $\epsilon_\theta=0.0015$ and (c) $\epsilon_\theta=0.002$. Dimensionless units. 
        }
\end{figure}
These scenarios are explicitly verified by numerical simulations 
for a helix consisting of $N=400$ sites, Fig.~\ref{fig10}.
When the torsional rigidity is zero ( $\epsilon_\theta=0$ ), and non-valent 
interactions dominate the elastic response, 
the stretching is two-phase. The distribution of longitudinal extension
$\Delta l_n$ along the chain completely matches the expectation based 
on the shape of $E(\Delta L)$ function, that is for 
$\Delta L\le 0.07$ the chain is stretched uniformly, while for $0.07\le\Delta L\le 0.165$ non-uniform stretching is observed.
The terminal regions of the chain are in a strongly stretched state, 
while the central region is stretched weakly [Fig.~\ref{fig10} (a)]. 
The transition boundary between the states  
is clearly seen in Fig.~\ref{fig7} (b). As the helix is extended further, 
the weakly stretched central region shrinks and 
vanishes when $\Delta L =0.17$. Beyond that point 
the helix is stretched uniformly. The domain of the 
non-uniform stretching regime decreases for higher torsional rigidity 
$\epsilon_\theta=0.0015$ [Fig.~\ref{fig10} (b)], and at even higher values 
$\epsilon_\theta >0.002$ only uniform stretching is observed [Fig.~\ref{fig10} (c)].

Thus,  if the torsional rigidity is small enough,
 stretching of the $\alpha$-helix proceeds via two-phase scenario with a 
typical plateau region where the tension remains constant. The 
scenario is confirmed by all-atom molecular dynamics 
simulations \cite{Zegarra2009} and experiments \cite{Schwaiger2002, Afrin2009}.
The root cause of the non-uniform stretching in this case is 
the typical form of the hydrogen bond potential (\ref{f4}), which has an inflection point. In contrast, polymer helices that 
have no hydrogen bonds, such as Polytetrafluoroethylene (PTFE) helix, are 
expected to stretch uniformly, without force-extension plateaux.

\subsection{Stretching of the DNA double helix}
The coarse-grained model \cite{Savin2011,kikot2011new} 
used here to simulate stretching of the DNA double 
helix (dsDNA) is semi-atomistic: each 
nucleotide consists of six united atom particle  -- three for the sugar-phosphate backbone and three for the nucleobase, see section \ref{s4} and appendix \ref{sd}.    

\begin{figure}[thb]
\includegraphics[width=1\linewidth]{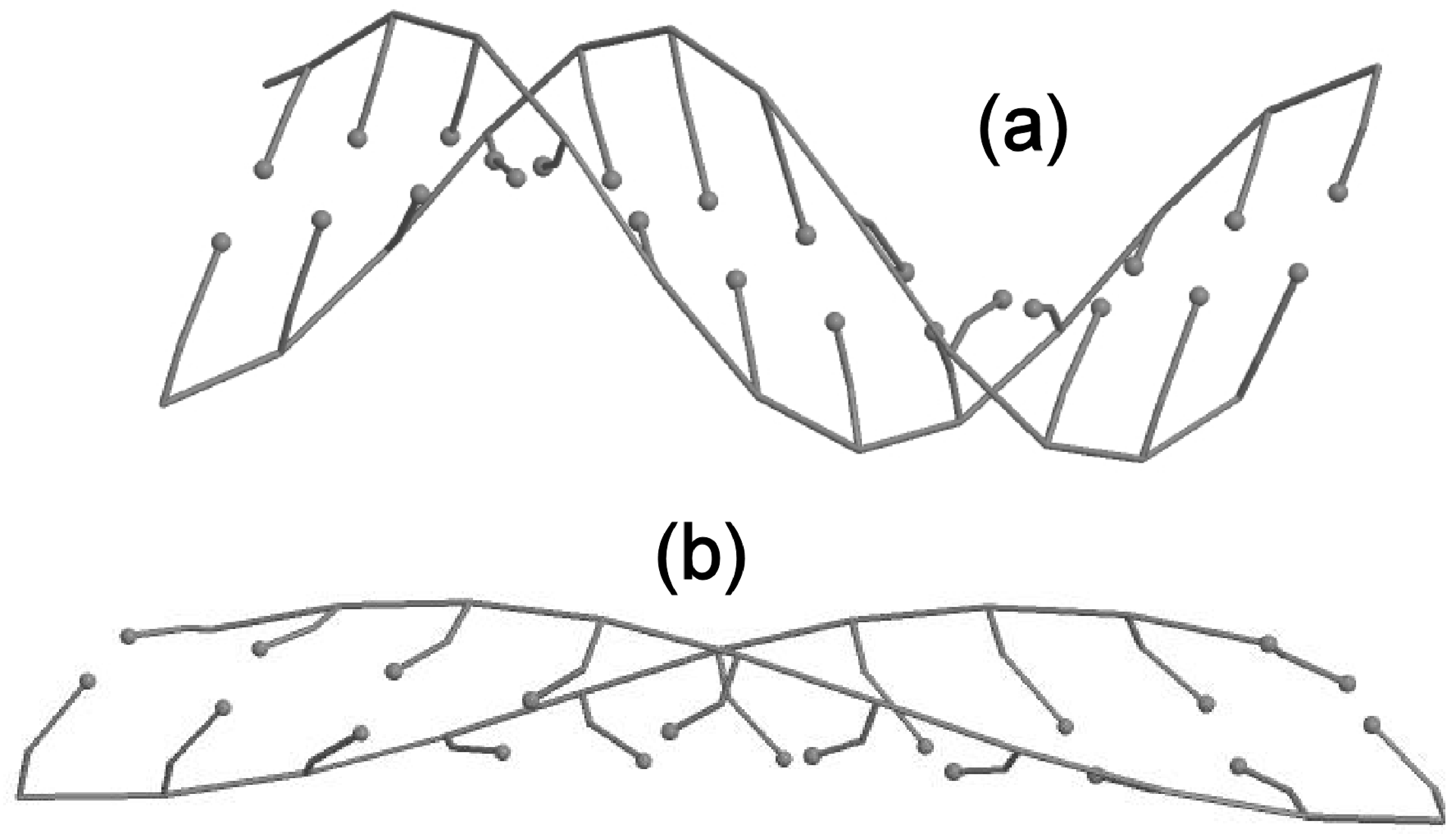}
\caption{\label{fig13_ira}
Structure of dsDNA fragment in (a) weakly stretched  (longitudinal step is $l=l_a=3.75$\AA) state and (b) strongly stretched  (longitudinal step is  $l=l_b=6.15$\AA) state. Minimum energy (ground) states are shown.  
        }
\end{figure}

\begin{figure}[tb]
\includegraphics[trim=0 20 0 20, angle=0, width=1\linewidth]{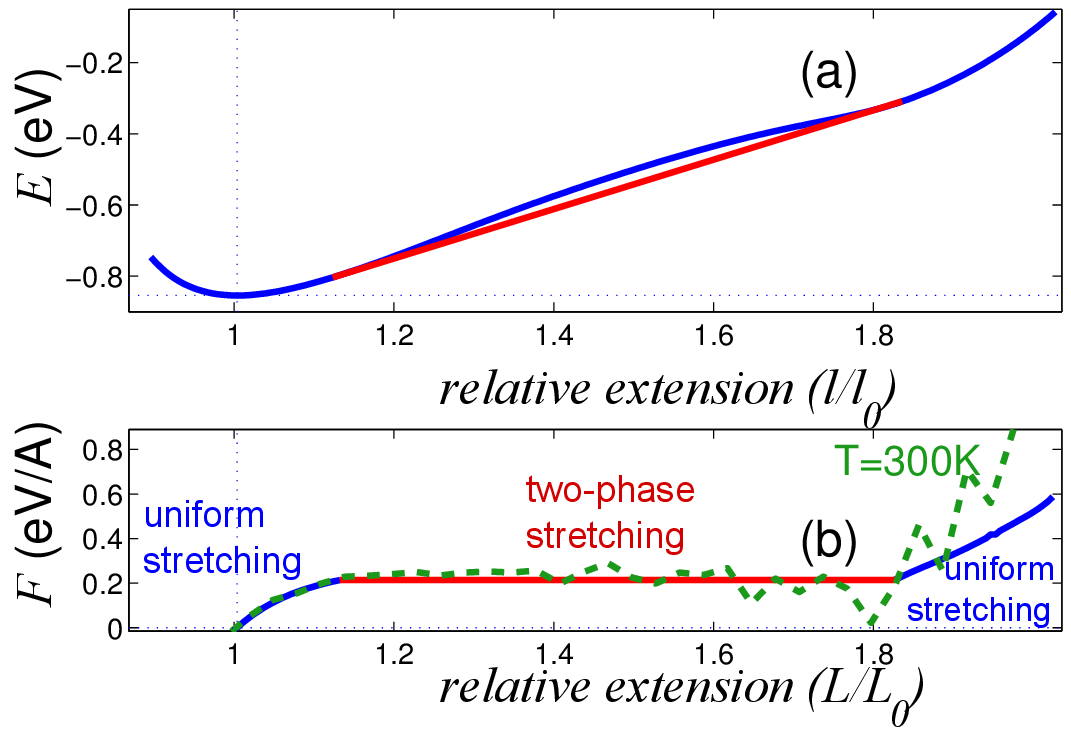}
\caption{\label{fig15}
(a)   Blue: Effective energy $E(l/l_0)$ per base pair of   
extended poly(A)-poly(T) DNA, ground state. 
Red: the convex hull of $E(l/l_0)$.  
(b) The tension as a function of the relative chain extension. Blue and red: 
in the ground state; green dashed: at T=300K.  
}  
\end{figure}

The corresponding effective
site (base pair) energy $E(l)$ as a function of relative site extension 
is shown in Fig.~\ref{fig15} (a).
To facilitate direct comparison with experiment, 
the energy is taken to depend on 
the relative extension $l/l_0$ instead of absolute deviation $\Delta l$ from 
equilibrium base-pair length $l_0$;  $l_0=3.352$A  calculated 
within our model agrees with the experimental value for B-DNA.   
The function $E(l/l_0)$ is non-convex between points $l_a/l_0=1.12$, $l_b/l_0=1.84$, 
its convex hull is shown by red line 2 in Fig.~\ref{fig15} (a). 
Thus, when the mean relative site extension $L/l_0$ of a stretched 
dsDNA fragment is between the above two values, 
a part of the double helix is in the weakly extended state with the 
longitudinal step $l_a$, while the rest of the base-pairs are 
in the strongly extended state with longitudinal step $l_b$, 
Fig.~\ref{fig13_ira}.  The corresponding 
force-extension diagram of the chain is shown in Fig.~\ref{fig15} (b). 

\begin{figure}[tb]
\includegraphics[width=1\linewidth]{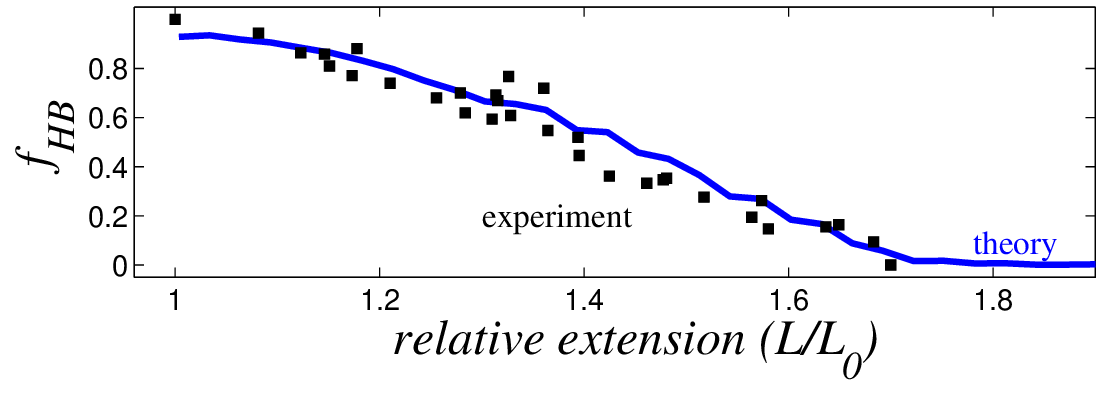}
\caption{\label{figHB_AT}
Fraction of remaining hydrogen bonds as a function of relative dsDNA 
extension. Blue line: simulation at 300K, black squares: experiment(Ref. \cite{vanMameren2009}).  
        }
\end{figure}

The  proposed non-uniform stretching
mechanism, so far explored without taking into account 
thermal fluctuations, holds at room temperature:  
only the range of the dsDNA over-stretching 
plateau increases slightly, Fig.~\ref{fig15}. 
Critically,  the room temperature 
value of the tension at the 
plateau coincides with the value obtained from the analysis of 
the minimum energy (ground) states described above.  
As the model chain stretches at room temperature, 
thermal fluctuations cause the WC hydrogen bonds to break,
as expected from  experiment, Fig.~\ref{figHB_AT}

In the plateau regime, the
DNA double-helix consists of two fractions:
a slightly stretched helix with
hydrogen bonds intact,
and a strongly stretched helix with
some hydrogen bonds broken (Fig.~\ref{fig14}).
This exact behavior is
observed in all-atom simulations \cite{Harris2005, Li2009}.

An important question arises whether the breaking of the 
hydrogen bonds between complementary bases 
is necessary \cite{Rouzina2001_1, Rouzina2001_2, Williams2004, McCauley2008} for the 
observed over-stretching plateau, or is the unzipping of the helix simply the 
consequence of the helix stretching? Our analysis clearly shows that  
WC bond breaking is not necessary for the appearance of the 
over-stretching plateau. First, 
the two-phase stretching behavior of dsDNA and the
force-extension plateau [Fig.~\ref{fig15} (b)] with virtually 
the same characteristics exist in  
the absence of thermal fluctuations when hydrogen bonds are
only weakened in the strongly stretched regime, but not yet broken. 
Second, in a computational experiment in which the bond 
strength is artificially doubled to prevent breaking of WC bonds 
at room temperature, we find virtually the same plateau. 
This explains the somewhat puzzling result of a 
recent single-molecule experiment in which torsionally
relaxed DNA exhibited the same over-stretching plateau when its unzipping
was inhibited \cite{Paik2011}. 
        In the case of the double-stranded DNA, it is mainly 
the base-stacking deformations 
that give the effective stretching energy its 
non-convex shape that is ultimately responsible 
for the onset of the two-phase stretching with the characteristic plateau, 
Fig.~\ref{fig13}.   Mathematically, base stacking is described 
by a combination of power law functions -- Coulomb and Lennard-Jones potentials -- that give it the non-convex shape. 

The plateau in dsDNA force-extension diagram was 
observed previously in single molecule stretching
experiments \cite{Smith, Cluzel}; the plateau was found in the range of 
(relative) extensions $1.1<L/l_0<1.7$. This agrees well with our result 
$1.12<L/l_0 < 1.84$, Fig.~\ref{fig15} (b).
The value of the plateau transition force within our model 
is 0.2 eV/A = 320 pN, which is somewhat higher than experimental
estimates. A typical value of the force is often 
reported to be about $65\div 70$ pN \cite{Smith, Cluzel, vanMameren2009};
however one should keep in mind that the experimental value was obtained
in the type of experiment when the DNA strands are pulled 
by the same type (3$'$ or 5$'$) end, 
while the other two ends remain free.
In contrast, the simulations reported here correspond 
to uniform pulling by all four ends. Experimentally, larger 
values of the plateau tension were reported under the uniform 
pulling scenario: $105\div 120$pN  \cite{Clausen-Schaumann2000, vanMameren2009}.
 Although this value is still less than half of the 320 pN  
predicted by our model, we note that the above experiment used a
specific, non-homogeneous sequence. Many 
DNA properties are strongly sequence-dependent:
for example, pure poly(dG-dC) and poly(dA-dT) DNA sequences 
yield tension values at the plateau that differ by a factor of two \cite{Clausen-Schaumann2000}. 
Thus, only semi-quantitative agreement
of our homogeneous (poly-A) model with the above experiments may be expected.
Reported differences between earlier estimates based on all-atom 
room temperature Molecular Dynamics simulations and experimental
values are also of the same order \cite{Cluzel, Lebrun1996}; 
therefore numerical agreement we have obtained with experiment 
can be considered as reasonable.

\section{Conclusion\label{s3}}

While all polymers behave very similar to each other 
under weak tension where their 
elastic properties are entropic in nature and are virtually independent of the structure of the monomers, 
striking differences are observed in experiments when stronger forces are applied and short-scale details start to dominate. We show that these differences can 
be explained by a very general mechanism based on convexity of the (effective) 
deformation energy function of individual monomers. We demonstrate that 
when this energy is a convex function of 
the extension, the chain stretching is single-phase uniform, 
without a plateau  in the force-extension diagram. The 
scenario is realized in polymers such as polyethylene whose structure 
is supported by strong covalent interactions. In contract, when the 
secondary structure of a polymer is mostly due to weak non-covalent 
interactions, the deformation function may become non-convex, 
leading to two-phase  stretching: a part of the chain is stretched weakly, 
while the other is stretched strongly. In this regime, 
extension of the whole chain proceeds by increasing of 
the fraction of the strongly stretched sites, 
so the tension remains constant. The force-stretching diagram 
has the characteristic plateau seen in experiment. 
Examples include $\alpha$-helix polypeptide,
and DNA double helix,
consistent with earlier observations based on all-atom molecular dynamics 
simulations and previous single molecules experiments.  
        We illustrate the general mechanism by numerical 
simulations based on realistic coarse-grained models and atomistic 
potentials of several polymers, 
from planar "zig-zag", to the more complex helix and B-DNA. Numerical
modeling is in complete agreement with the general mechanism, and 
in acceptable agreement with experiment. 

        The main strengths of the proposed theory are its complete generality 
and direct connection to microscopic structure of the monomers.
Our framework applies to any polymer in the strong deformation 
regime where short-scale
details dominate -- to the best of our knowledge, no such 
universal description was available before. Although division of the polymer 
into two stretching phases was discussed earlier in the context 
of phenomenological models  \cite{Storm03, Rief98}, 
the second equilibrium state was assumed to exist 
and to be known {\it a priori}. Within our framework, no assumptions 
of multiple stable equilibrium states \cite{Schwaiger2002,Rief98} or 
additional kinetic arguments are necessary \cite{Rief98}.



\section{Methods\label{s4}}

\subsection{Computing the force-extension diagram}
The effective site deformation energy 
function $E(\Delta l)$ (Fig.~\ref{fig1})
is found by minimizing the 
total potential energy $H$ of the chain under 
the constraint that each site is stretched by the same amount 
$\Delta l$: 
$E(\Delta l) = N^{-1} \min\limits_{\Delta l_i = \Delta l} \{H\}$.  
This definition of $E(\Delta l)$ includes contributions from 
both short-range interactions within each site and short- and long- range
interactions between the sites.

Next, we consider the same chain  of $N \gg 1$ effective sites, but 
without the $\Delta l_i = \Delta l$ constraint, 
that is with the possibility of non-uniform extension.  
The dependence of the mean 
chain energy $E_{NU}(\Delta L)$ upon its mean 
longitudinal extension $\Delta L = N^{-1} \sum_{i=1}^N \Delta l_i$ is found 
by minimizing the total energy of the chain $H$ 
under the condition of fixed 
total deformation $\sum_i^N \Delta l_i$: 
$E_{NU}(\Delta L) = N^{-1} \min\limits_{ N^{-1} \sum_{i=1}^N \Delta l_i = \Delta L} \{H \}$. 
The resulting  function $E_{NU}(\Delta L)$ is the 
convex hull of the effective site deformation energy $E(\Delta l)$ 
(see appendix \ref{sa} for details ).  
Unless otherwise specified, the extension force (tension) is obtained as
$F={ dE_{NU}(\Delta L)}/{d \Delta L}$.
No torsional constraints are imposed in any case. Unless otherwise stated, 
polymer chain is  modeled as quasi-one-dimensional crystal. 

\subsection{2D plane zigzag}
The 2D zigzag chain, Fig.~\ref{fig2}, is specified by the distance between its neighboring 
sites $\rho_0$ and the zigzag angle $\phi_0$ (equilibrium longitudinal step is 
$l_0=\rho_0\sin(\phi_0/2)$). We consider dimensionless zigzag model,
details of its parameters values are given in Refs. \cite{Zolotaryuk96,Manevitch97,Savin99} 
and the appendix \ref{sb}. 

The chain potential energy is given by:
\begin{equation}
H=\sum_n\{V(\rho_n)+\epsilon_\phi U(\phi_n)+W(r_n)\}.
\label{f1}
\end{equation}
where $V(\rho_n)$ is the valent bond energy between neighboring atoms 
$n$  and $(n+1)$ separated by distance $\rho_n$. 
\begin{equation}
V(\rho)=\frac12K(\rho-\rho_0)^2,
\label{f2}
\end{equation}
with the bond rigidity $K=2$.
The $\epsilon_\phi U(\phi_n)$ term is the deformation energy 
of the angle between atoms $(n-1)$, $n$  and  $(n+1)$, where 
$\epsilon_\phi\ge 0$ is the angle deformation stiffness.
\begin{equation}
U(\phi)= (\cos\phi-\cos\phi_0)^2.
\label{f3}
\end{equation}
The last term $W(r_n)$ corresponds to weak non-valent interaction between atoms 
$n$ and $(n+2)$ (next-nearest neighbors) separated by $r_n$; its form is 
typical for non-valent
interactions; it may be used for description of hydrogen 
bonds and van der Waals interactions alike.  
\begin{equation}
W(r)=\epsilon_{hb}\left[\left(\frac{r_0-d}{r-d}\right)^6-1\right]^2,
\label{f4}
\end{equation}
where $\epsilon_{hb}\ge 0$ is the interaction energy, $r_0=2h_x=1.633$ is 
the equilibrium length, $d=0.5$ is the diameter of the inner hard core. 
The balance between valent and non-valent interactions is varied 
by changing the angle stiffness $\epsilon_\phi$, while 
keeping other interactions fixed.

\subsection{$\alpha$-helix}
The model we describe here, Fig.~\ref{fig7} (a), 
is similar to the ones \cite{Christiansen97,Savin2000}
used previously for analysis of ultrasonic soliton motion.
The equilibrium atomic helix coordinates:
\begin{equation}
{\bf R}_n^0=(R_0\cos(n\varphi_0),R_0\sin(n\varphi_0),nl_0),
\label{f7}
\end{equation}
with $n=0,\pm1,\pm2,...$ being the atom number, 
$R_0$ -- helical radius, $\varphi_0$ and $l_0$
-- the angular and longitudinal helix period.

The chain potential energy is
\begin{equation}
H=\sum_n\{V(\rho_n)+\epsilon_\phi U(\phi_n)+ \epsilon_\theta Z(\theta_n)+W(r_n)\}.
\label{f8}
\end{equation}
 The term $V(\rho_n)$ gives the energy of interaction between neighbor
sites $n$ and $(n+1)$, where $\rho_n$ is the distance between them.
The angle deformation  energy is described by $\epsilon_\phi U(\phi_n)$,
where $\phi_n$ is the angle between sites $(n-1)$, $n$, and $(n+1)$ 
(the vertex is on site $n$).
The third term  $\epsilon_\theta Z(\theta_n)$ is the energy 
of the torsional deformation (rotation) around $n$-th bond:
\begin{equation}
\epsilon_\theta Z(\theta)=\epsilon_\theta\left(\cos\theta-\cos\theta_0\right)^2,
\label{f9}
\end{equation}
where $\epsilon_\theta\ge 0$ is torsional rigidity;
different values of $\epsilon_\theta$ are considered while keeping other
interactions fixed.
The function $W(r_n)$
is the energy of the $n$-th hydrogen bond connecting
sited $n$ and $(n+3)$.
We consider dimensionless model of the helix, see Appendixes for specific parameters. 
The bond deformation energy is described by the potential (\ref{f2}) with the rigidity $K=10$. Hydrogen bond energy is given by eq. (\ref{f4}), 
angle deformation energy is described by (\ref{f3}). 

\subsection{DNA double helix}
The potential energy of the double helix consists of 4 terms:
\begin{equation}
 H=E_{A} + E_{B} + E_{st} + E_{hb}^*.
 \label{DNA_hamiltonian}
 \end{equation}
The first two terms describe deformation energy of 
complementary strands A and B, respectively, 
within the 12CG coarse-grained model \cite{Savin2011}. 
These terms 
include essentially the same energetic contributions as in the case of the
$alpha$-helix: internal energy (bond stretching, angle bending and torsion twisting) plus non-valent interaction between the grains within the same strand. 
The last two terms are                                 
non-valent interactions:  $E_{st}$ -- between two neighboring base pairs, 
and $E_{hb}^*$ -- between 
two complementary bases (within the same base-pair), including hydrogen bonds. 

        Within the framework of our 
coarse-grained model \cite{Savin2011}, 
the nitrogen bases are treated most accurately, at all-atom level. 
WC hydrogen bonds and stacking interactions are modeled 
via Coulomb and van der Waals potentials taken 
from current all-atom AMBER  \cite{amb1} force-field 
widely used to model nucleic acids.  
To make the computations
feasible, solvent effects 
are treated via the so-called implicit 
solvation model \cite{Honig1995} 
at the generalized Born level often used in all-atom simulations of 
DNA \cite{VTsui2000}. Within the model, water is 
treated as a continuum with the (room-temperature) 
dielectric and hydrophobic properties of water;
screening effect of salt ions is also taken into account. Hydrogen bonding with 
the solvent is present, albeit in an ``average" sense. The 
balance between solute-solute and solute-solvent h-bond strength is 
controlled by adjusting $E_{hb}$ -- 
the interactions between complementary bases taken 
from the all-atom AMBER explicit solvent force-field \cite{amb1}.
Here we use $E_{hb}^* = c_0 E_{hb}$, with $c_0=0.4$, which leads to 
quantitative agreement with experiment, Fig.~\ref{figHB_AT}.    
Further details of the calculation
can be found in Ref.  \cite{Savin2011} and SI.
To avoid sequence-dependence issues that do not affect the basic physics, 
we consider homogeneous poly(A)-poly(T) sequence.  

The tension at $T=300K$, Fig.~\ref{fig15} (b), is obtained as $F=d\langle H \rangle /dL$, 
where $\langle\cdot \rangle$ denotes ensemble averaging 
over a Molecular Dynamics trajectory. The simulation employed a 500 base-pair 
poly(A)-poly(T) fragment in the 12CG coarse-grained representation, 
see Appendixes for details. The same trajectory was 
used to obtain results in Fig.~\ref{figHB_AT}.

\begin{acknowledgments}
This research was supported by RFBR (grant no. 08-04-
91118-a), CRDF (grant no. RUB2-2920-MO-07), and, in part, NIH GM076121 
to A.V.O. The authors thank
Erwin J.G. Peterman for a helpful discussion and  
providing experimental data points in Fig.~\ref{figHB_AT}.

\end{acknowledgments}

\appendix
\section{Minimal energy conformation of the polymer chain is determined by 
convex properties of the effective deformation energy curve 
of individual monomer site\label{sa}} 
Consider a linear polymer chain of $N \gg 1$ identical sites (monomeric units).
Each site $i$ is extended by $\Delta l_i$, with the 
corresponding deformation energy $E(\Delta l_i)$. 
When pulled by both ends, the equilibrium 
 energy of the entire chain  can be found by solving 
the constrained (conditional) minimization problem:
 
 $$NE_{NU}(\Delta L)=\min\limits_{\Delta l_1 + \dots + \Delta l_n=N\Delta L}  \{ E(\Delta l_1) + \dots + E(\Delta l_n) \}  $$
 
 where $N \Delta L$ is the total extension of the chain.  
 Here $\Delta L$ denotes the mean per site extension of the chain. 
Using the 
 method of Lagrange multipliers,  the problem is converted to the unconditional minimization problem over (N+1)  variables: 
 $$\min \{ E(\Delta l_1) + \dots + E(\Delta l_n)-\lambda (\Delta l_1 + \dots + \Delta l_n - N\Delta L)  \}.$$
 
 Differentiating  with respect to $\Delta l_i$ yields $E^\prime(\Delta l_i) - \lambda$ that should equal zero for each $i$. Derivative with respect to $\lambda$ also equals zero; this condition yields the original constraint. Excluding $\lambda$ from all of the equations, we obtain the following system of $N$ equations:
\begin{eqnarray}
E^\prime(\Delta l_1)=E^\prime(\Delta l_2)=\dots=E^\prime(\Delta l_N) 
\label{eq:lagrange1} \\
\Delta l_1 + \dots + \Delta l_n=N\Delta L
\nonumber
\end{eqnarray} 

When the function $E(\Delta l)$ is convex, its derivative is a monotonically increasing function, 
and the condition  \ref{eq:lagrange1}  can be satisfied only when 
extensions of all the sites are equal to each other: 
$\Delta l_a = \dots = \Delta l_n= \Delta l = \Delta L$, that
is when the chain is extended uniformly.  
In contrast, when  the energy function $E(\Delta l)$ is non-convex, 
it is possible for its derivative (tangent) to have the same value 
at two distinct points $\Delta l_a$ and $\Delta l_b$, 
see Fig.~\ref{fig1}. In this case, some $\Delta l_i$ (for simplicity, $i=1,\dots, pN$) are equal to $\Delta l_b$, while the rest of $\Delta l_i$ ($i=pN+1, \dots, N$) are equal to $\Delta l_a$, $\Delta l_a < \Delta l_b$. 
 If $N \gg 1$ (technically, if $N \rightarrow \infty$), one can always find such $p$,  $0 < p < 1$, 
that the constraint $\Delta L=p\Delta l_b + (1-p)\Delta l_a$ is satisfied. 
The deformation energy in this case of non-uniform stretching is equal to $pNE(\Delta l_b)+(1-p)NE(\Delta l_a)=N(pE(\Delta l_b)+(1-p)E(\Delta l_a))$, which  
is a linear function of $\Delta L$ that connects  
points $(\Delta l_a, E(\Delta l_a)),(\Delta l_b,E(\Delta l_b))$. This 
linear function is the convex hull of $E(\Delta l)$,  see Fig 1 in the main 
text. By definition of non-convex function, 
$pE(\Delta l_b)+(1-p)E(\Delta l_a) < E( p\Delta l_b + (1-p)\Delta l_a ) = E(\Delta L)$, 
which proves that the 
two-phase extension is energetically preferred 
relative to the uniform extension 
in this case. As the chain is stretched further, and $\Delta L$ increases, $p$ 
increases accordingly so that $\Delta L=p\Delta l_b + (1-p)\Delta l_a$ 
is satisfied.  The extension of the chain via the change in the fraction 
of sites extended by $\Delta l_b$ can continue until $p=1$ at $\Delta L = \Delta l_b$. 

The above reasoning does not take into account phase boundary effects. However, these, as well as end effects, are negligible as long as the polymer chain 
is long, $N gg 1$, which is always the case experimentally.   
Our numerical calculations give the same results supporting our general conclusions.  
        
In what follows we consider three microscopic polymer 
models in detail: 2D zigzag, $\alpha$-helix, and double-stranded DNA. 
Since we are interested in the regime where polymer extension
approaches its contour length, entropic contributions (key in weak stretching ) are neglected.   

\section{2D Zigzag\label{sb}}
Consider a dimensionless 2D model of zigzag chain shown in Fig.~\ref{fig2}. Such a chain can be considered as a quasi-one-dimensional crystal with the elementary cell being two neighboring sites; that is each cell can be obtained from the previous one by translation along the x-axes.
The potential energy of the zigzag chain is given by the Eq. (\ref{f1}). 
The equilibrium bond length is
$\rho_0=1$, zigzag angle  is $\phi_0=\arccos(-1/3)=109.47^\circ$ (so that equilibrium longitudinal step is $l_0=0.8165$). 
The non-valent interaction is given by $W(r)=\epsilon_{hb}\left[\left(\frac{r_0-d}{r-d}\right)^6-1\right]^2$, where $\epsilon_{hb}\ge 0$ is the interaction energy, $r_0=2h_x=1.633$ is 
the equilibrium length, $d=0.5$ is the diameter of the inner hard core. 
For simplicity, but without loss of generality, we set the equilibrium length
$r_0$ to correspond to the equilibrium value of the angle $\theta_0$. 
The non-valent interaction energy coefficient was fixed at 
$\epsilon_{hb}=0.0178$ (so that $W^{\prime\prime}(r_0)=72\epsilon_{hb}/(r_0-d)^2=1$).
\begin{figure}[tb]
\includegraphics[width=1\linewidth]{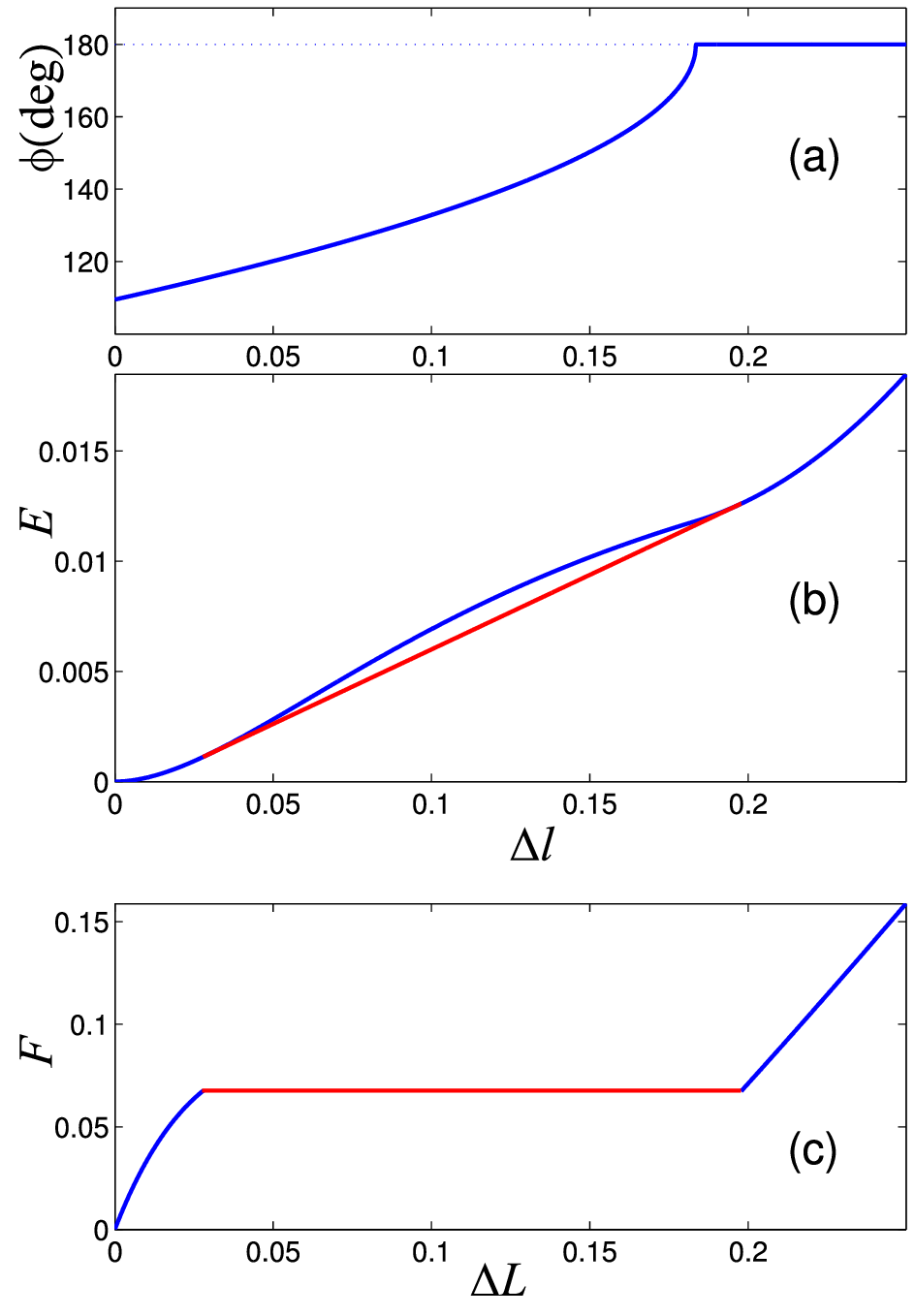}
\caption{\label{fig3_full}
2D zigzag. The dependence of (a) angle $\phi$, (b) effective site deformation energy $E$ of the zigzag upon  extension $\Delta l$ of its sites  and (c) the tension as 
 a function of the mean site extension.
Angle deformation stiffness is $\epsilon_\phi=0$, which leads to the two-phase
stretching regime. The red line in b) is the 
convex hull of $E(l)$.  Dimensionless units.
}
\end{figure}

Let the x-axis be along the chain, and the y-axis in the 
perpendicular direction, Fig.~\ref{fig2}. The longitudinal (x-) step $l$
and transverse (y-) step $h$ for site $n$ are shown in Fig.~\ref{fig2}.
To find the uniform extension energy that is energy of  a single 
zigzag site extended by $\Delta l>0$ along the x-axis, 
one has to solve the minimization problem over transverse step  $h$
\begin{equation}
E(\l)=V(\rho)+\epsilon_\phi U(\phi)+W(r)\rightarrow\min:h
\label{fa2}
\end{equation}
with fixed value of  the longitudinal step  $l_0+\Delta l$.
Note that the  distance between next-nearest neighbors 
is  $r=2(l_0+\Delta l)$, and the value of the angle
 $\phi$ is defined by the bond length  $\rho$.
Solving the minimization problem  (\ref{fa2}) yields  
the energy (per one site) a uniformly stretched zigzag as a function of  
the relative longitudinal extension $\Delta l$. 
Conjugate gradient is used to find the minimum. 

\begin{figure}[tb]
\includegraphics[width=1\linewidth]{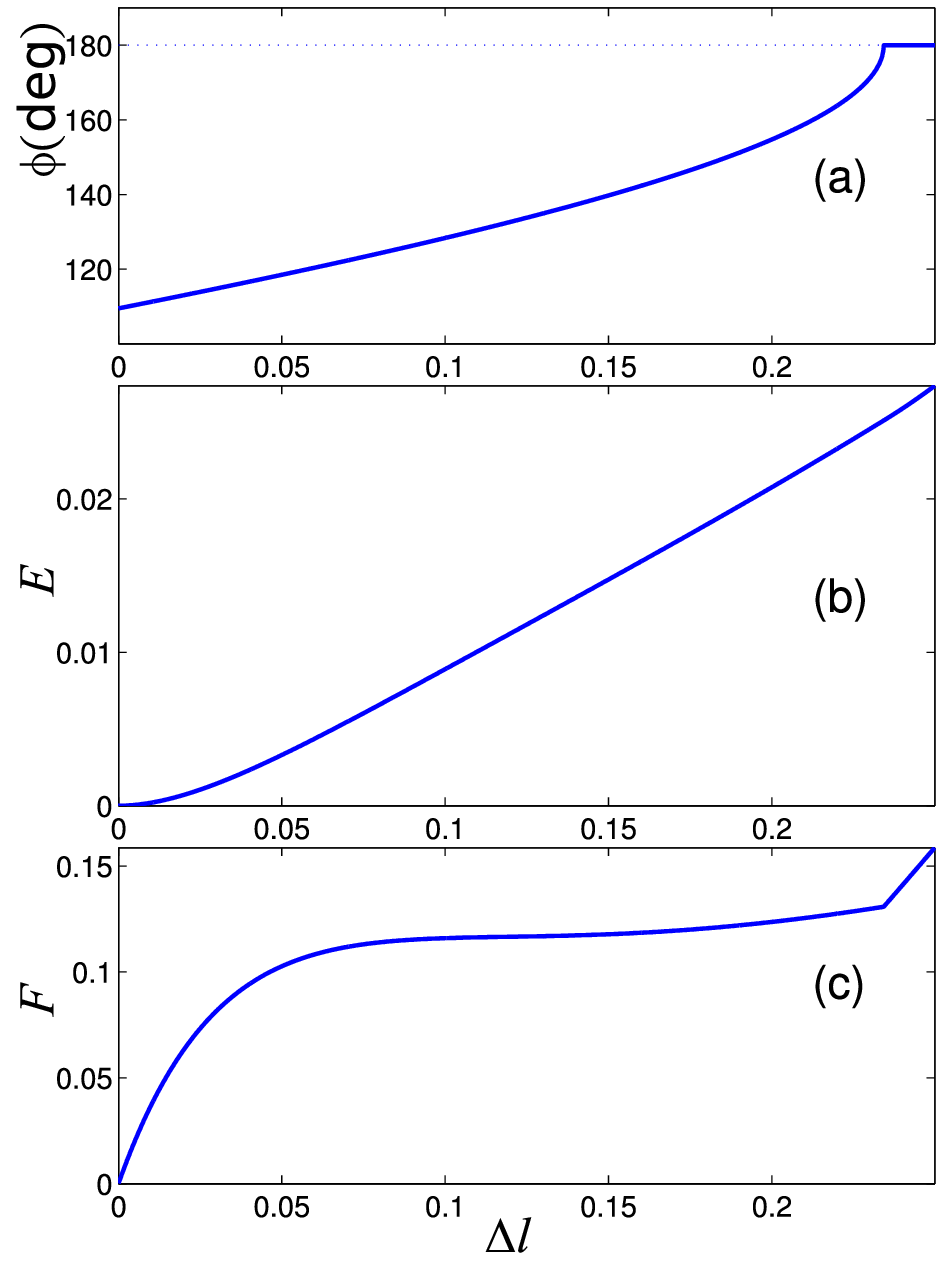}
\caption{\label{fig4_full}
2D zigzag. The dependence of (a) angle $\phi$, (b) effective site deformation energy $E$ and (c) the tension as a function of the site extension. 
Angle deformation stiffness is $\epsilon_\phi=0.02$; non-valent interactions
dominate the elastic response and so only uniform stretching is possible.
Dimensionless units.
        }
\end{figure}

The 
angle $\phi$, energy and tension force as functions of 
the extension are shown in 
Fig.~\ref{fig3_full}, \ref{fig4_full} for different values of the coefficient $\epsilon_\phi$. This coefficient controls the relative contribution of  a valent interaction (angle deformation energy) to the elastic response of the chain. 

To find the energy of the extended (generally non-uniformly) chain 
for each value of the mean (per site) extension
$\Delta L\ge 0$, one has to solve the minimization problem:
\begin{equation}
\sum_{n=1}^{N-1}V(\rho_n)+\sum_{n=2}^{N-1}\epsilon_\phi U(\phi_n)+\sum_{n=1}^{N-2}W(r_n)\rightarrow
\min_{\{x_n,y_n\}_{n=1}^N}
\label{fa3}
\end{equation}
with the condition of fixed ends:
$$
x_1\equiv -\frac12(N-1)\Delta L,~~ x_N=(N-1)l_0+\frac12(N-1)\Delta L.
$$

Conjugate gradient is used to find the minimum. The initial condition is 
chosen to be that of a uniformly stretched chain. 
It should be mentioned that in what follows only the x-coordinate is fixed, 
all other coordinates  are allowed to change freely.
If $\{ x_n,y_n\}_{n=1}^N$ is a solution of the minimization problem (\ref{fa3}),
then the distribution of the longitudinal extension in the chain is given by the function
$\Delta l_n=x_{n+1}-x_n-l_0$, and the distribution of the transverse step 
is given by the function $h_n=|y_{n+1}-y_n|$.
\begin{figure}[tb]
\includegraphics[width=1.\linewidth]{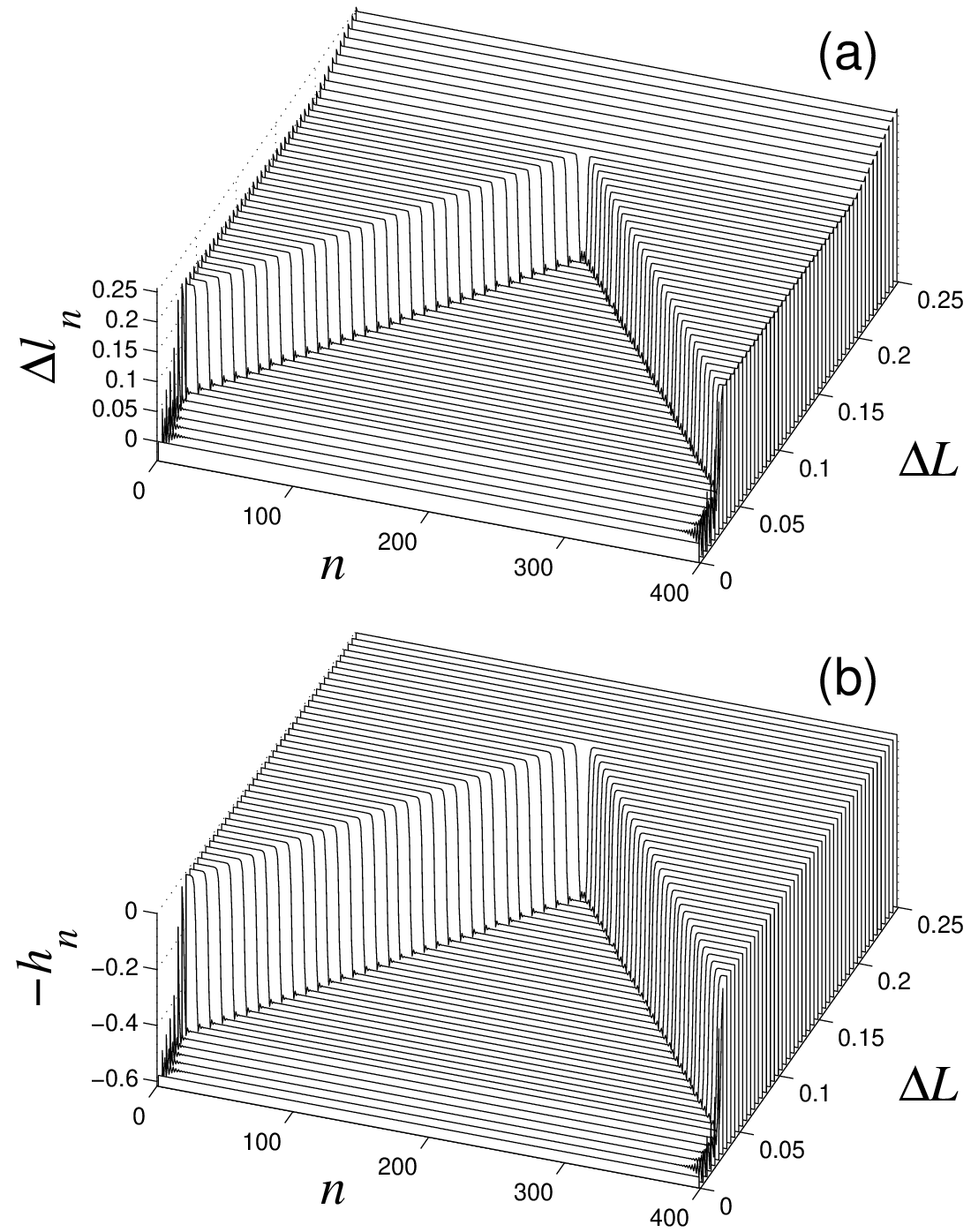}
\caption{\label{fig5}
The two-phase stretching in 2D zigzag chain. Shown is the dependence of site extension $\Delta l_n$ and transverse step  $h_n$
upon the site number $n$ and the mean (per site) extension $\Delta L$. 
A  chain with $N=400$ sites (atoms) with fixed ends. 
Angle deformation stiffness is $\epsilon_\phi=0$. Dimensionless units.
        }
\end{figure}

\begin{figure}[tb]
\begin{center}
\includegraphics[width=1.\linewidth]{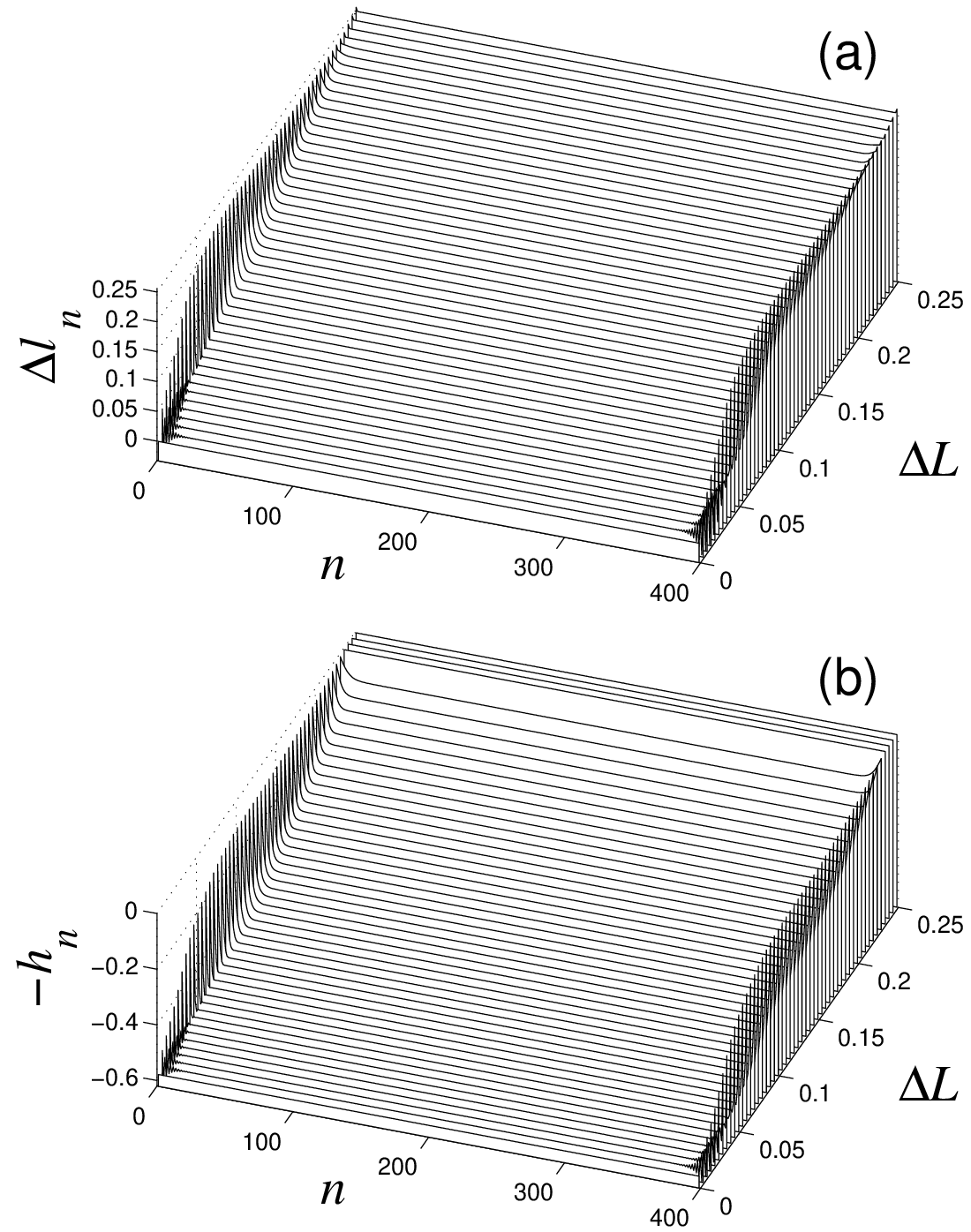}
\end{center}
\caption{\label{fig6}
The uniform stretching of the 2D zigzag chain. Shown is the dependence of site extension $\Delta l_n$ and transverse step  $h_n$
upon the site number $n$ and the mean (per site) extension $\Delta L$. 
A  chain with $N=400$ sites (atoms) with fixed ends.
Angle deformation stiffness is $\epsilon_\phi=0.02$. Dimensionless units.
        }
\end{figure}

One can see from Fig.~\ref{fig5} that when $\epsilon_\phi=0$ 
the distribution of the longitudinal step extension 
$\Delta l_n$ and the transverse step $h_n$ along the chain is in agreement with the  prediction that can be made based on convexity of the function 
$E(\Delta l)$. That is  when $0\le\Delta l\le 0.037$ and $\Delta l\ge 0.2056$, 
uniform stretching takes place,
while when  $0.037<\Delta l<0.2056$ non-uniform one is observed.
Here, the end sites of the zigzag are in the strongly extended 
state with  extension $\Delta l_n=0.2056$,
while the central part is in weakly stretched state with  extension $\Delta l_n=0.037$.
Gradual transition from one state to the other is observed along the domain 
boundary.
As the chain extends, the size of the weakly stretched part in the center 
monotonously decreases and vanishes at $\Delta l=0.2056$. 
At even higher tension, the zigzag is stretched uniformly.
The two-phase stretching scenario persists for as 
long as  $0 \le \epsilon_\phi \le 0.015$. The range of  extensions 
where the scenario is realized gradually shrinks as the contribution 
of the valent bond potential grows.  
Once $\epsilon_\phi>0.015$,  
all the sites are stretched uniformly, Fig.~\ref{fig6}. 
One can see from Fig.~\ref{fig3_full} that the 
strongly extended state corresponds to the completely stretched out zigzag
(angle $\phi$=180$^0$).

\section{Alpha-helix\label{sc}}
Consider a 3D molecular chain 
corresponding to an ideal \cite{Khokhlov1994} 
$\alpha$-helix.
The equilibrium atomic helix coordinates:
$$
{\bf R}_n^0=(R_0\cos(n\varphi_0),R_0\sin(n\varphi_0),nl_0),
$$
with $n=0,\pm1,\pm2,...$ being the atom number, 
$R_0$ -- helical radius, $\varphi_0$ and $l_0$
-- the angular and longitudinal helix period.
For the sake of simplicity we consider dimensionless model of the helix [see Fig.~\ref{fig7} (a)], where
the (dimensionless) helix radius  is $R_0=0.4919$, 
angular step is $\varphi_0=100^\circ$,
longitudinal step is $l_0 = 0.6572$ \cite{Christiansen97}.
Such a helix can be treated as a quasi-one-dimensional crystal; 
that is each site can be obtained from the previous one by the appropriate 
translation along longitudinal axes and rotation around the same axis.
\begin{figure}[t]
\includegraphics[width=1.\linewidth]{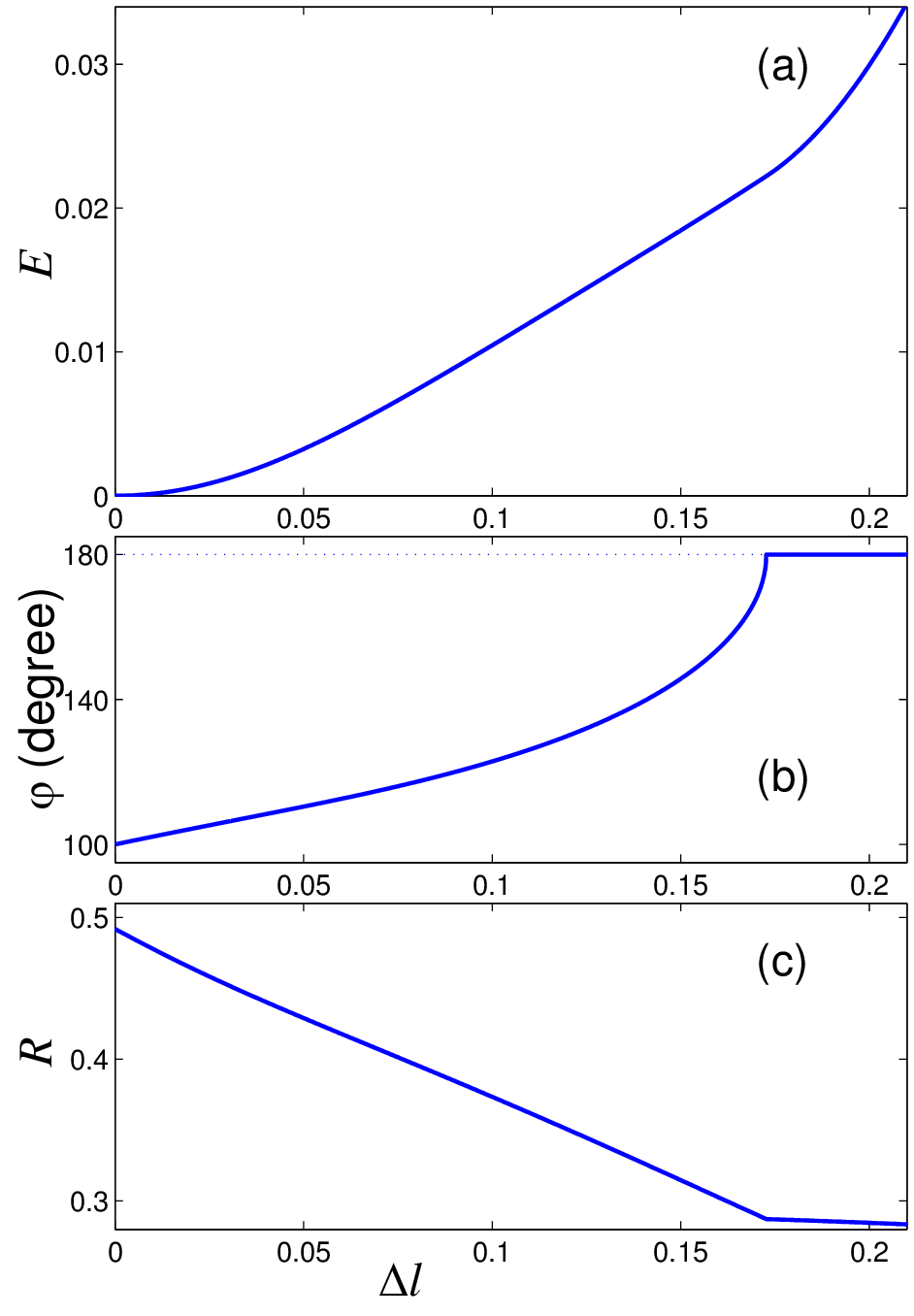}
\caption{\label{fig9_full}
Dependence of the effective site energy $E$, (b) angular step $\varphi$ and (c) radius $R$
of the helix from upon its relative (uniform) longitudinal site 
extension $\Delta l$.
Torsional rigidity is $\epsilon_\theta=0.002$. Dimensionless units.
        }
\end{figure}

The chain potential energy is
$$
H=\sum_n\{V(\rho_n)+\epsilon_\phi U(\phi_n)+ \epsilon_\theta Z(\theta_n)+W(r_n)\}.
$$
 The term $V(\rho_n)$ gives the energy of interaction between neighbor
sites $n$ and $(n+1)$, where $\rho_n$ is the distance between them.
Bond rigidity is $K=10$, 
the equilibrium bond length is $\rho_0=1$.
The angle deformation  energy is described by $\epsilon_\phi U(\phi_n)$,
where $\phi_n$ is the angle between sites $(n-1)$, $n$, and $(n+1)$ 
(the vertex is on site $n$). Equilibrium angle is $\phi_0=\arccos(-1/3)=109.47^\circ$,
coefficient $\epsilon_\phi=1$.
The specific form of these terms 
is described in ``Methods" section.
The third term  $\epsilon_\theta Z(\theta_n)$ is the energy 
of the torsional deformation (rotation) around $n$-th bond. 
$$
\epsilon_\theta Z(\theta)=\epsilon_\theta\left(\cos\theta-\cos\theta_0\right)^2,
$$
where $\epsilon_\theta\ge 0$ is the torsional rigidity;
different values of $\epsilon_\theta$ are considered while keeping other
interactions fixed. The equilibrium torsion angle is $\theta_0=\arccos(0.2395)$. 
The function $W(r_n)$
is the energy of the $n$-th hydrogen bond connecting
sited $n$ and $(n+3)$. It is given by the formula
$$
W(r)=\epsilon_{hb}\left[\left(\frac{r_0-d}{r-d}\right)^6-1\right]^2,
$$
The equilibrium hydrogen bond length is $r_0=2.0322$.
Other parameters of the hydrogen bonding potential are: 
$d=0.7$(inner core diameter), $\epsilon_{hb}=0.0246$, so that rigidity of non-valent interactions $W^{\prime\prime}(r_0)=72\epsilon_{hb}/(r_0-d)^2=1$. Equilibrium values of angles and distances correspond to 
specified values of helix radius $R_0$, angle step $\varphi_0$ and longitudinal step $l_0$ of the helix in ground state.

        To determine the dependence of the effective helix site energy $E$
(per one step) upon the relative uniform 
extension $\Delta l$, we 
solve the following conditional minimization problem: 
$$
E(R, \varphi, l_0+\Delta l)=V(\rho)+\epsilon_\phi U(\phi)+\epsilon_\theta Z(\theta)+W(r)\rightarrow \min_{R, \varphi}
$$
with the fixed value of longitudinal step $l_0+\Delta l$. Here $R$ is the helix 
radius, and $\varphi$ is its angular step, see section ``Methods". 
Conjugate gradient is used to find the minimum. 

In general, to find variable extensions $ \Delta l_n$ of each helix site in the case of non-uniform stretching, 
we solve: 
\begin{eqnarray}
\sum_{n=1}^{N-1}V(\rho_n)+\sum_{n=2}^{N-1}\epsilon_\phi U(\phi_n)+\sum_{n=2}^{N-2}\epsilon_\theta Z(\theta_n)
+\sum_{n=1}^{N-3}W(r_n) \nonumber\\
\rightarrow \min: \{(R_n,\varphi_n,l_0+\Delta l_n)\}_{n=1}^N
\nonumber
\end{eqnarray}
with the fixed ends condition:
$$
l_1\equiv -\frac12(N-1)\Delta L,~~ l_N\equiv (N-1)l_0+\frac12(N-1)\Delta L.\
$$  Here, as before, $N\Delta L$ is the total extension of the chain of $N$ 
sites. 
Conjugate gradient is used to find the minimum. The initial condition is
chosen to be that of the uniformly stretched helix. 

\section{dsDNA\label{sd}}

\subsection{Model details} 
\begin{figure}[t]
\begin{center}
\includegraphics[angle=90, width=.58\linewidth]{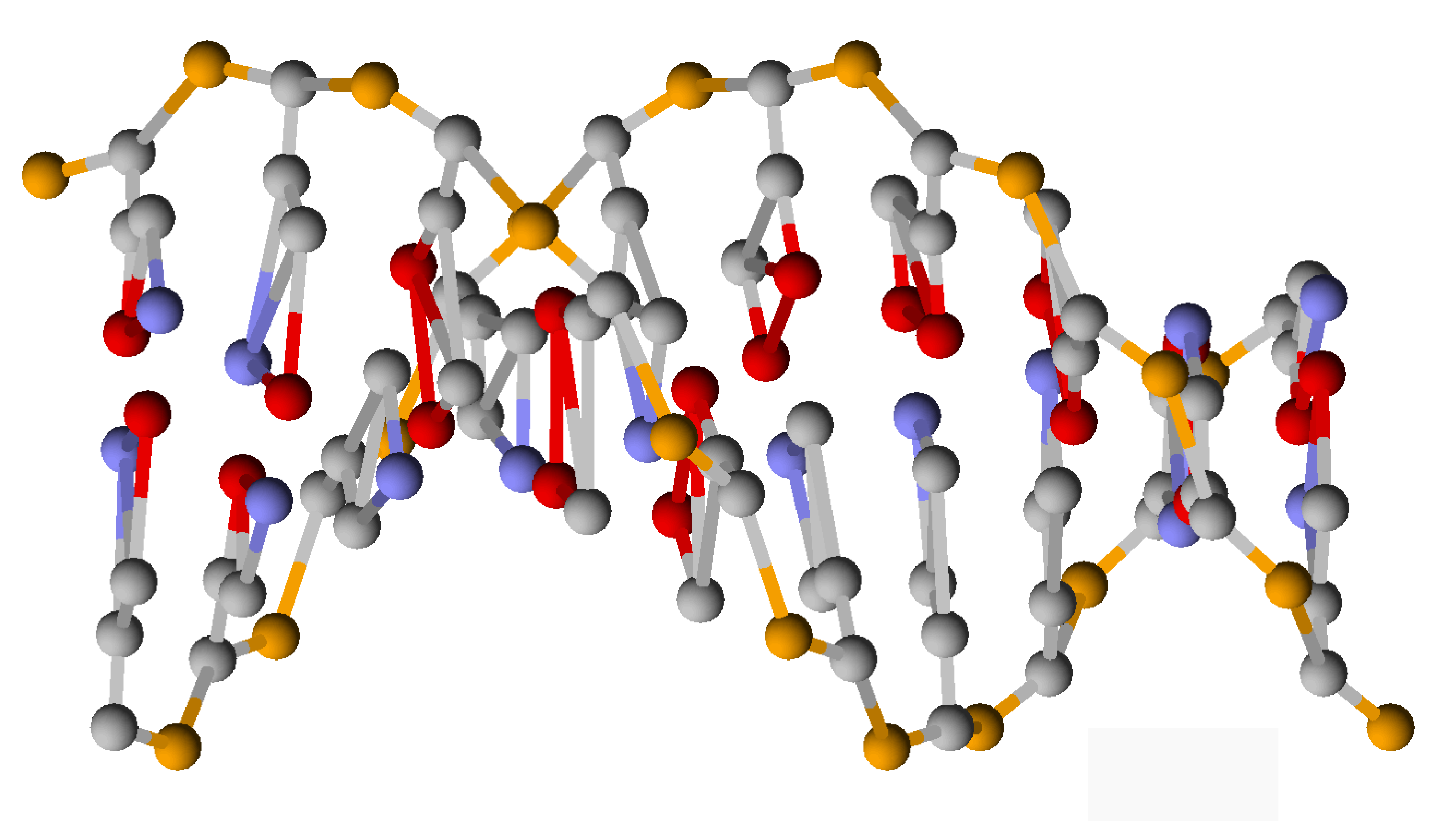}
\end{center}
\caption{\label{fig01c}\protect
A DNA fragment in the coarse-grained representation \cite{Savin2011}
used here. Each base pair (site) is modeled by 12 united
atom particles (grains).}
\end{figure}

\begin{figure}[tb]
\begin{center}
\includegraphics[width=0.7\linewidth]{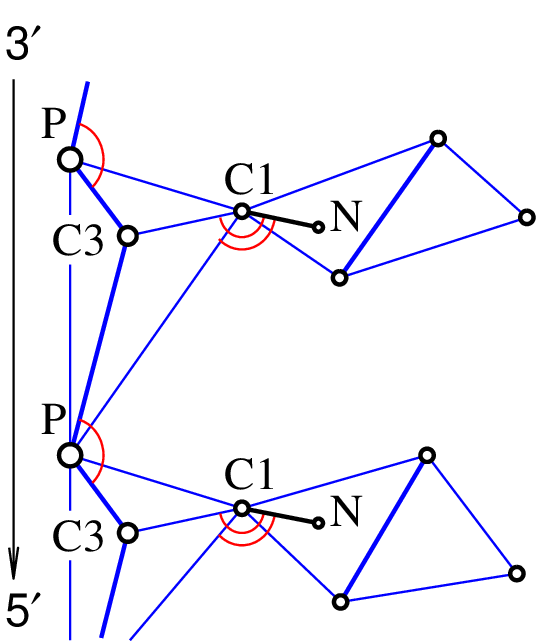}
\end{center}
\caption{\label{fig11}
United atom particles (grains) involved in the valent interactions 
in the 12CG coarse-grained DNA model. 
Blue lines denote  valent (harmonic) bonds,
red arcs mark valent angles, bold lines are axes of rotation in the 
torsional potentials. 
The circles marked as N stand for atoms N9 in A and G bases,
and N1 in T and C bases (no grain is situated on these atoms, their
coordinates are calculated directly from positions of the 
base grains as detailed in Ref.  \cite{Savin2011}).}
\end{figure}

The 12CG model of the DNA double-helix used in this work is 
shown in Figs.~\ref{fig01c} and ~\ref{fig11}. To provide additional 
information we switch to atomic units. 

The total potential energy 
of the system has the following form:
\begin{equation}
H=[E_v+E_b+E_a+E_t+E_{el}+E_{vdW}]+E_{hb}^{*} +E_{st}.
\label{f12}
\end{equation}

The terms in the brackets  describe the deformation energy of both strands. 
In the main part, a short-hand notation was used to avoid 
unnecessary details. For example, the energy 
$E_A$ of  strand "A", which appears 
in the main part, equals the bracketed terms above in which  
only the untied atoms from strand "A" are retained.  
The terms  $E_v$, $E_a$, $E_t$ correspond to valent bond, angle and torsion  
deformation energy respectively.
These potentials have a common form: bond deformation energy is calculated 
as
$$
U_{\alpha \beta}(r)= \frac12 K_{\alpha \beta}(r-R_{\alpha \beta})^2,
$$
with the rigidity coefficients  $K_{\alpha \beta}$ 
and equilibrium values  $R_{\alpha \beta}$
being different for different type grains.
Angle deformation energy has the form
$$
U_a(\phi)=\epsilon_a(\cos\phi-\cos\phi_a)^2,
$$
where values of the coefficient $\epsilon_a$ and the 
equilibrium angle differ for different angle types.
Torsion deformation energy is described by the potential
$$
U_{t}=\epsilon_{t}(1-\cos(\theta-\theta_0)),
$$
where values of the coefficient $\epsilon_t$ and the equilibrium torsion angle $\theta_0$ differ for different torsions.

Rotational axis of the torsional potential are shown in Fig.~\ref{fig11}.
The third term  $E_b$ in the energy function (\ref{f12}) describes deformation energy of a nitrogen base.
The nitrogen base is a rather rigid chemical structure modeled here 
by rigid harmonic potentials which
keep four points near one plain: grain $C1$ 
and the three points on each nucleobase. 

The next two terms $E_{el}$, $E_{vdW}$ describe electrostatic and van der Waals interactions
between backbone grains. Solvent is treated implicitly, 
via the Generalized Born (GB) model \cite{Still1990, Onufriev1102}. 
The methodology has been used to model free DNA in solution \cite{VTsui2001,sorin03}, 
binding between proteins and nucleic 
acids \cite{DeCastro:2002:J-Mol-Recognit:12382239,Allawi03,Chocholousova06}, 
conformational changes such as the  
$A \rightarrow B$ transition \cite{VTsui2000}, as well as for 
exploring dynamics of long DNA fragments \cite{JZmuda06}. 
The GB model approximates solvation energy of two interacting charges by the following formula
originally proposed by Still et al. \cite{Still1990}
$$
  \Delta G_{solv} \approx
   - \frac{1}{2} \left( 1 - {{1}\over{{\epsilon}_{out}}} \right)
   \sum_{ij} \frac{q_i q_j}{f(r_{ij}, R_i, R_j)},
$$
where $\epsilon_{out}$ is the dielectric constant of water,
$r_{ij}$ is the distance between atoms $i$ and $j$,
$q_i$ is the partial charge of atom $i$,
$R_i$ is the so-called \emph{effective Born radius} of atom $i$,
and $f = {\Big[r_{ij}^2 + R_i R_j \exp({-r_{ij}^2/{4 R_i R_j})} \Big]}^{1\over2}$.
\begin{figure}[tb]
\includegraphics[width=1.0\linewidth]{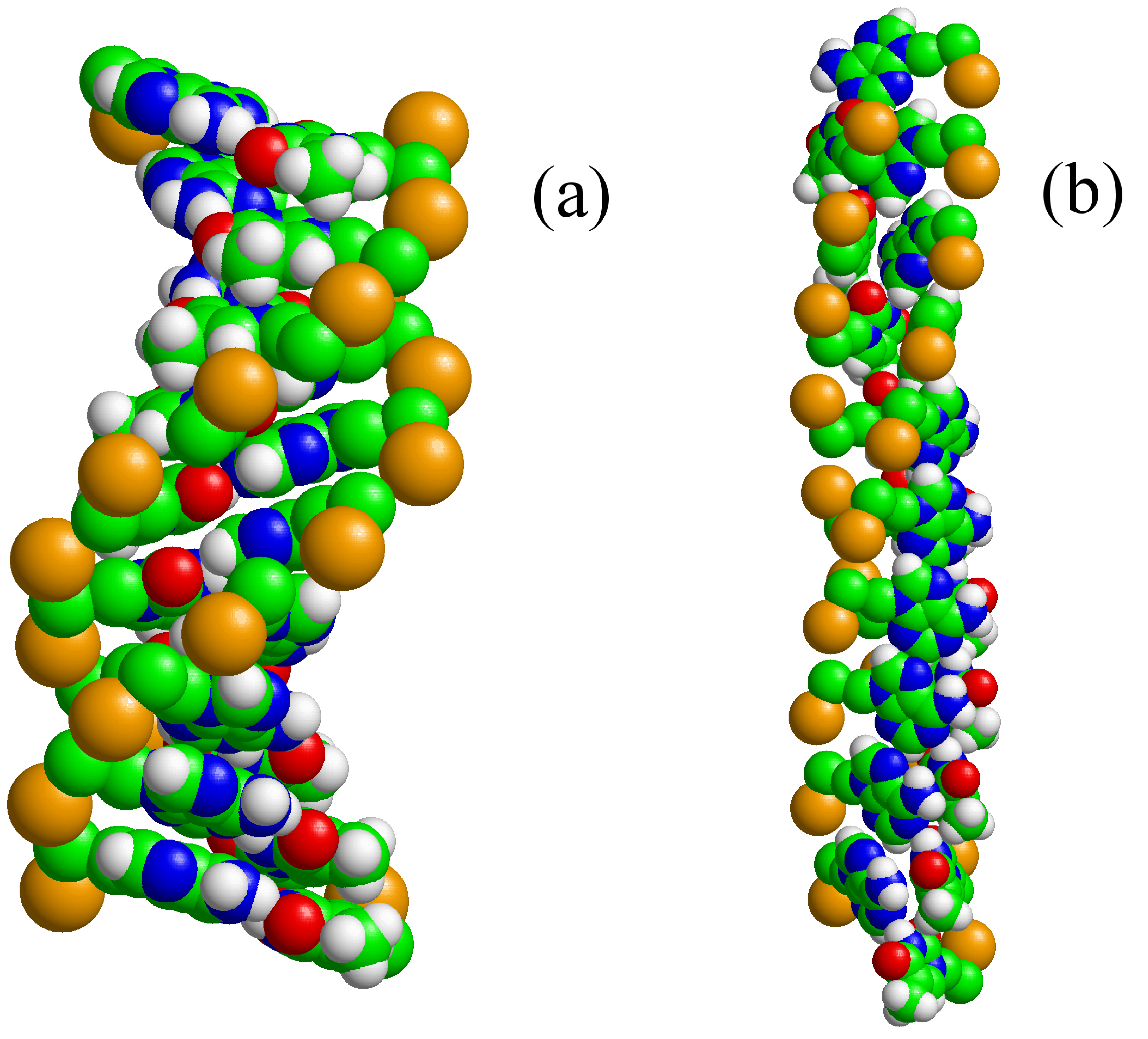}
\caption{\label{fig13}
Schematic view of the minimum energy (ground) state of (a) weakly stretched  (longitudinal step is $l=l_a=3.75$\AA) and
(b) strongly stretched  (longitudinal step is  $l=l_b=6.15$\AA) poly(A)-poly(T) 
DNA double helix.
        }
\end{figure}

The empirical function $f$ is designed to interpolate between the limits
of large $r_{ij} \gg \sqrt{R_i R_j}$  where the Coulomb law applies, and
the opposite limit where the two atomic spheres fuse into one, restoring the
famous Born formula for solvation energy of a single ion.
The effective Born radius of an atom represents its degree of burial
within the low dielectric interior of the molecule: the further away is
the atom from the solvent, the larger is its effective radius.
In our model, we assume constant effective Born radii which we calculate
once from the first principles \cite{Onufriev1102} for the B-form DNA. 
The screening effects
of monovalent salt are introduced approximately, at the Debye-Huckel level by
substitution
$$
1 - {\epsilon_{out}}^{-1} \rightarrow
1 - {\epsilon_{out}}^{-1} \exp( -0.73 \kappa f).
$$
The 0.73 pre-factor was found empirically to give the best agreement with the numerical
Poisson-Boltzmann (PB) treatment \cite{Case1999}. 
Here $\kappa$ is the Debye-Huckel screening parameter
$\kappa$[\AA$^{-1}$]$\approx 0.316 \sqrt{{\rm[salt][mol/L]}}$. Implementation
details in the context of the 12CG DNA coarse-grained model can be found 
in Ref.~\cite{Savin2011}.

The last two terms $E_{hb}^*$ and $E_{st}$ in Eq. (\ref{f12}) describe 
interactions between nitrogen bases (including
stacking and hydrogen bonds). Since nitrogen base is a rather rigid structure,
we can calculate coordinates of all the original atoms (corresponding to the 
all-atom representation) from positions of the three united atoms. This allows
us to utilize, directly, the 
all-atom AMBER \cite{amb1} Coulomb and van der Waals 
potentials used to mimic hydrogen bonds and stacking \cite{Savin2011}.

Assuming no sequence variability along the strand, e.g. poly(A)-poly(T), 
such double helix can be considered as a quasi-one-dimensional crystal with the elementary cell being one nucleotide pair of the double helix. 
In the ground (minimum energy) state each successive nucleotide pair is obtained from its predecessor by translation
along the z-axis by step $l$   
followed by 
a rotation around the same axis through helical step  $\Delta\phi$. 
(Here we use $l$ instead of $\Delta l=l-l_0$, because, unlike in the case 
of a ``simple" structure such as the 2D zigzag, the 
equilibrium value of the DNA longitudinal step $l_0$ in our model is not 
known {\it a priori}, 
and is obtained  by solving the corresponding
minimization problem.) 
Thus, the energy of the ground state is a function of 38 variables:
36 Cartesian coordinates of 12 grains in the first nucleotide pair,  and $\Delta\phi$, $l$.

To find the minimum energy (ground) state of the homogeneous (that is no sequence variability along the strand) double helix under tension,
we solve the following minimization problem over 37 variables
\begin{eqnarray}
H=(E_v+E_b+E_a+E_t+E_{el}+E_{vdW})+E_{hb}^{*} +E_{st}\nonumber\\
\rightarrow\min: \{ {\bf x_{j}}\}_{j=1}^{12},~\Delta\phi,~~
\nonumber
\end{eqnarray}
under the fixed value of the longitudinal step $l$. 
The summation is taken over only one base pair and neighboring base pairs are obtained 
from it by rotation and translation. Conjugate gradient is used to find the minimum; the initial condition corresponds to all-atom B-form DNA. 
The minimization yields the effective site energy (per one base pair)
$E(l)$ as a function of the longitudinal step $l$ -- see Fig.~\ref{fig12}. 
The energy minimum is reached for the 
longitudinal step $l_0=3.352$\AA,
which corresponds to the B-form of dsDNA.
\begin{figure}[tb]
\includegraphics[width=1\linewidth]{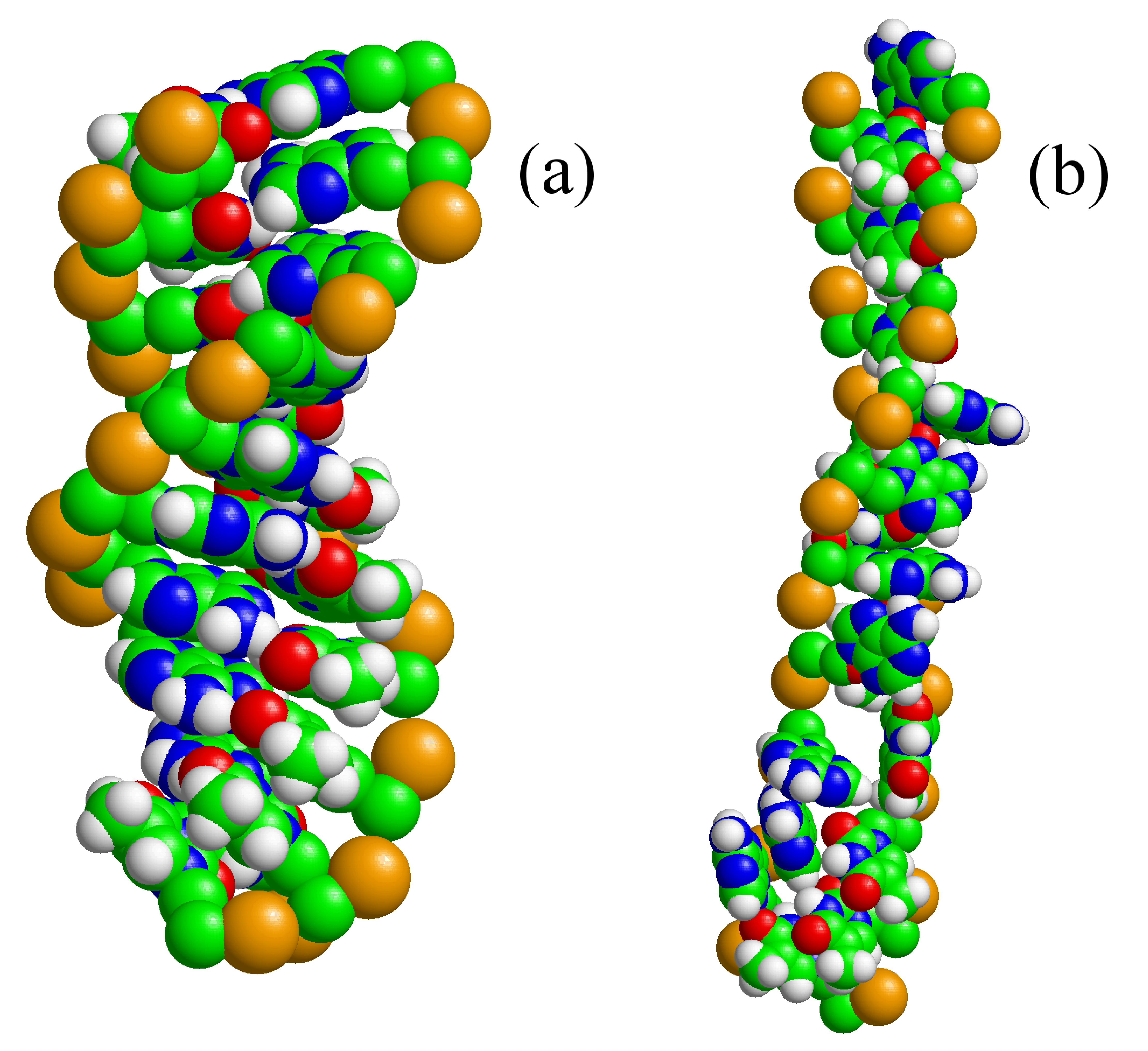}
\caption{\label{fig14}
Schematic view of a fragment of (a) weakly stretched  (averaged longitudinal step is $L=l_a=3.75$\AA) and
(b) strongly stretched (averaged longitudinal step is $L=l_b=6.15$\AA) poly(A)-poly(T) 
DNA double helix at $T=300$~K.
        }
\end{figure}

Since the effective site energy is non-convex, we predict 
two-phase stretching for dsDNA (see Figs. \ref{fig13}, \ref{fig14}) based on the general mechanism
described in  the main part. The DNA structure and its model potentials 
are most complex  among the three polymer models 
analyzed in this work. We have 
therefore chosen the DNA to further test our general predictions through 
molecular dynamics simulations at $300K$ (for the 2D zig-zag and the alpha-helix 
only purely mechanical stretching without thermal fluctuations 
was considered).

\subsection{Origins of the non-convex shape of the effective 
site deformation for double-stranded DNA } 

\begin{figure}[tb]
\includegraphics[width=1\linewidth]{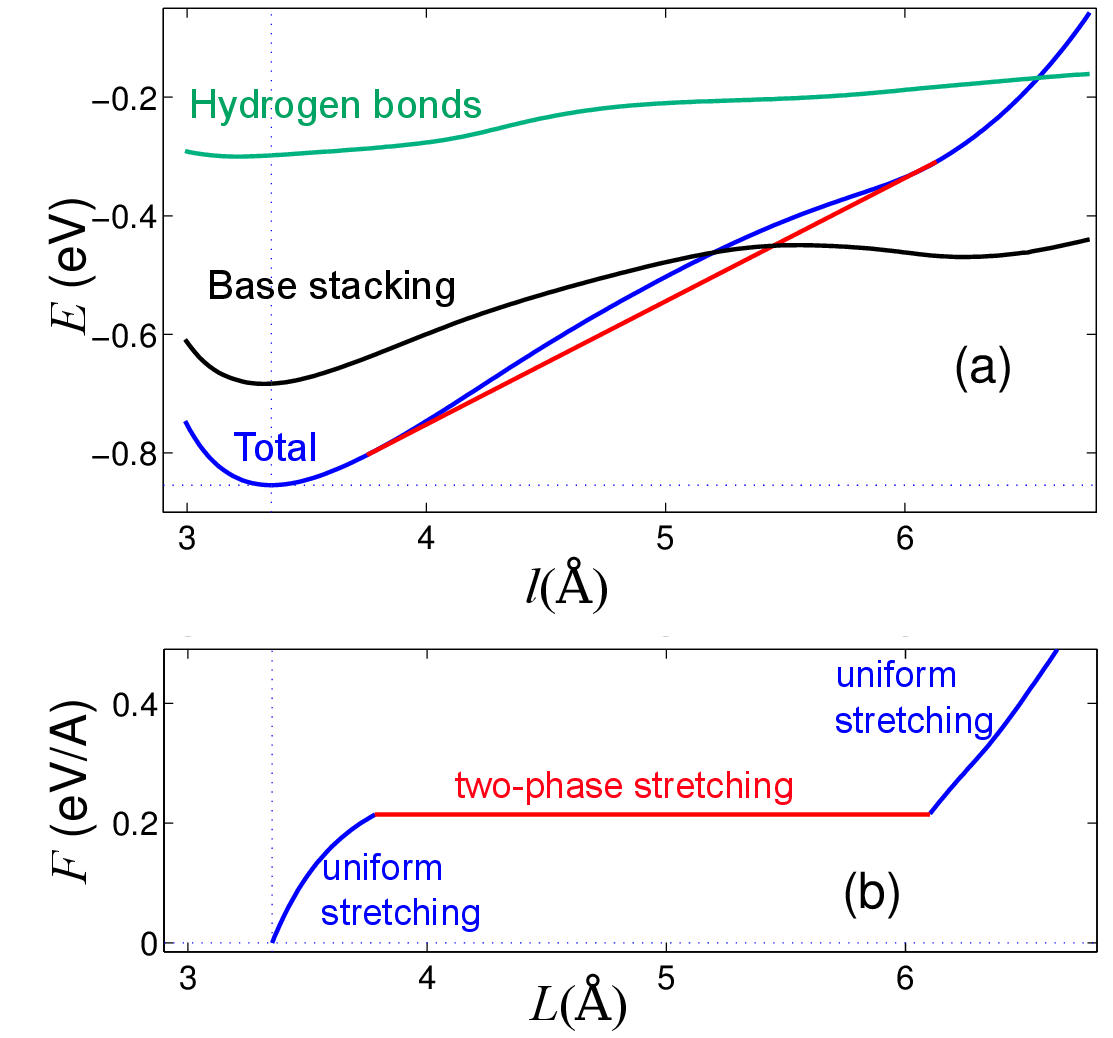}
\caption{\label{fig12}
(a) Effective energy $E(l)$ per base pair of extended
poly(A)-poly(T) DNA in ground state, hydrogen bond energy $E_{hb}^*$, and
neighbor base-pair stacking interaction energy $E_{st}$ as a function of  
longitudinal step $l$.
(b) The tension as a function of the average longitudinal step $L$.
}
\end{figure}

Variation of the base stacking and
hydrogen bond components of the effective site deformation
energy as a function of the chain extension for the double-helix DNA
is shown in Fig.~\ref{fig12} (a). In this computation, thermal
fluctuations are not considered. As the tension grows,
the base stacking weakens, the corresponding energy curve has a distinct
non-convex region. The hydrogen bonds also weaken, but do not break; the 
non-convex region on this curve is much less prominent. 

\subsection{Room temperature simulations of the DNA}

To bring the temperature of the molecule to the desired value $T=300K$,
we integrate over time the Langevin system of equations of motion:

$$
{\bf M}_n\ddot{\bf r}_n=-\partial H/\partial{\bf r}_n -\Gamma{\bf M}_n\dot{\bf r}_n+\Xi_n,
$$
where the index $n=1,2,...,N$ runs over all of the united 
atoms (grains), Figs. \ref{fig01c} and \ref{fig11}, 
$\Gamma=1/t_r$ is the Langevin collision frequency
with  $t_r=1$ ps being the corresponding particle  relaxation time,
${\bf M}_n$ is the mass of $n$-th united atom,
and $\Xi_n(t),{n=1}^{N}$ is a set of $N$ 3-dimensional vectors of independent Gaussian
distributed stochastic forces describing the interaction of $n$-th united atom
 with the thermostat with correlation functions
$$
\langle \Xi_n(t_1)\Xi_m(t_2)\rangle=2M_n\Gamma k_BT\delta_{nm}\delta(t_2-t_1).
$$
The initial conditions correspond to the
equilibrium state of the double helix.
\begin{figure}[tb]
\includegraphics[width=1.0\linewidth]{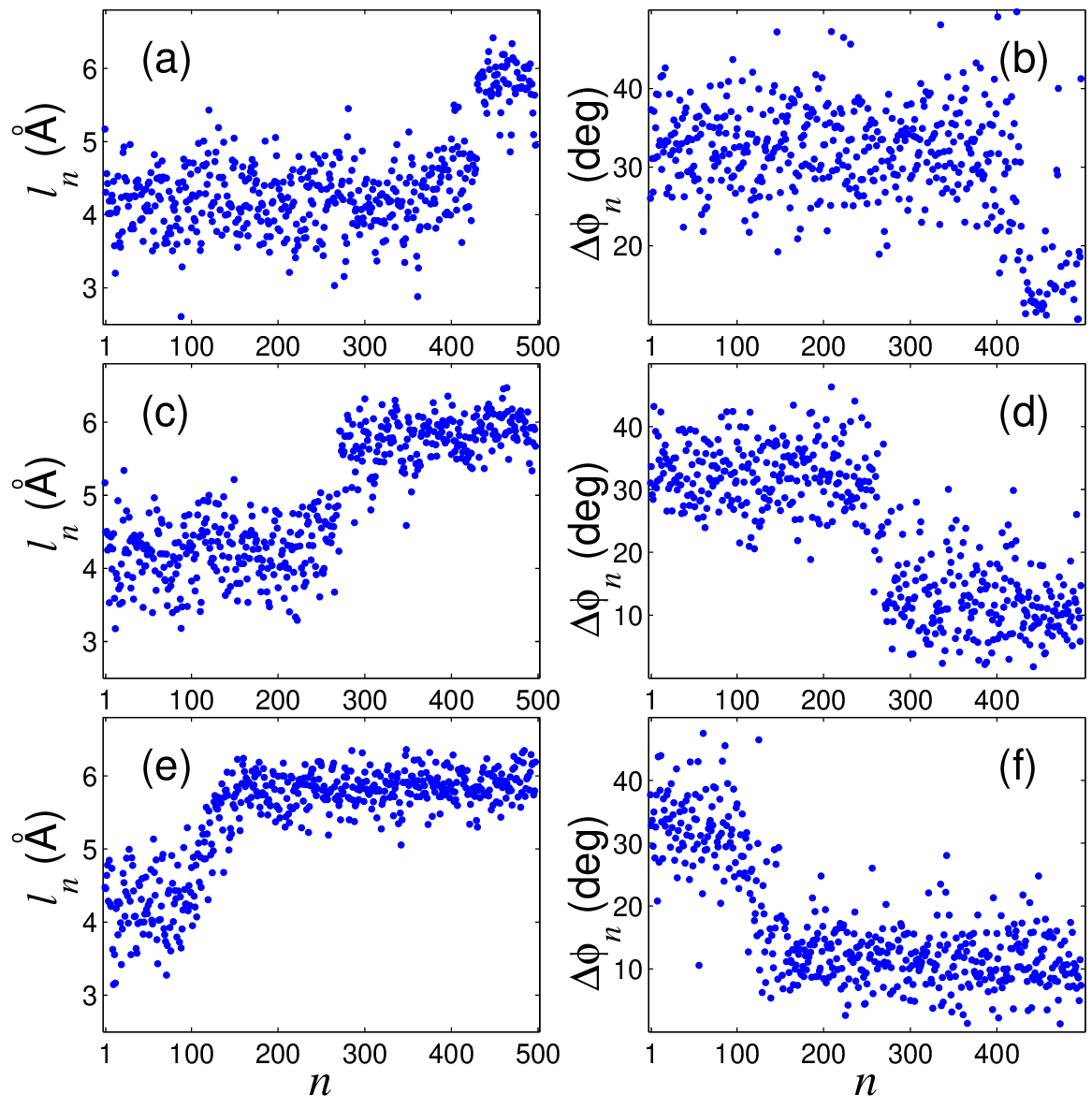}
\caption{\label{table1}
The two-phase stretching in dsDNA. Shown is the dependence of site length $l_n$ and angular step (twist) $\Delta\phi_n$ upon the base pair number $n$ along the DNA chain. A  500 b.p. poly(A)-poly(T) fragment was used in 
the simulation at 300K;  several values of the relative mean extension $L/l_0$ are tested; (a,b): $ L/l_0=1.33$; (c,d):$L/l_0=1.48$; (e,f):$L/l_0=1.63$.}
\end{figure}

Once the system is thermalized, the temperature is maintained at $T=300K$
and the trajectory continues for the desired simulation time. We use 
Verlet  integrator with the integration time step of 
$0.5$ fs.

Dependence of the dsDNA site length and angular step (twist) is shown in 
Fig.~\ref{table1}.
Twist was calculated using an {\it in-house} software based on the algorithms 
described in Ref.~\cite{Lu2003}. The length was 
calculated as the distance between neighbouring phosphorus atoms along the 
longitudinal axis (averaged over both strands).
\begin{figure}[tb]
\includegraphics[width=1\linewidth]{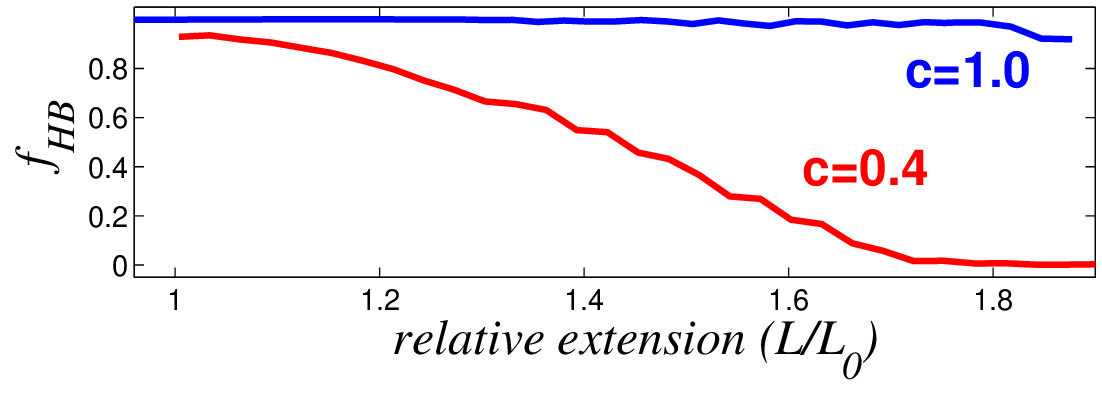}
\caption{\label{fig_cmp_fHB}
Fraction of remaining hydrogen bonds as a function of relative chain extension 
for dsDNA. Blue line: artificially strong WC bonds ($c_0=1.0$). 
Red line: regular strength WC bonds ($c_0=0.4$).  Simulation at $T=300K$ 
        }
\end{figure}
\begin{figure}[t]
\includegraphics[width=1\linewidth]{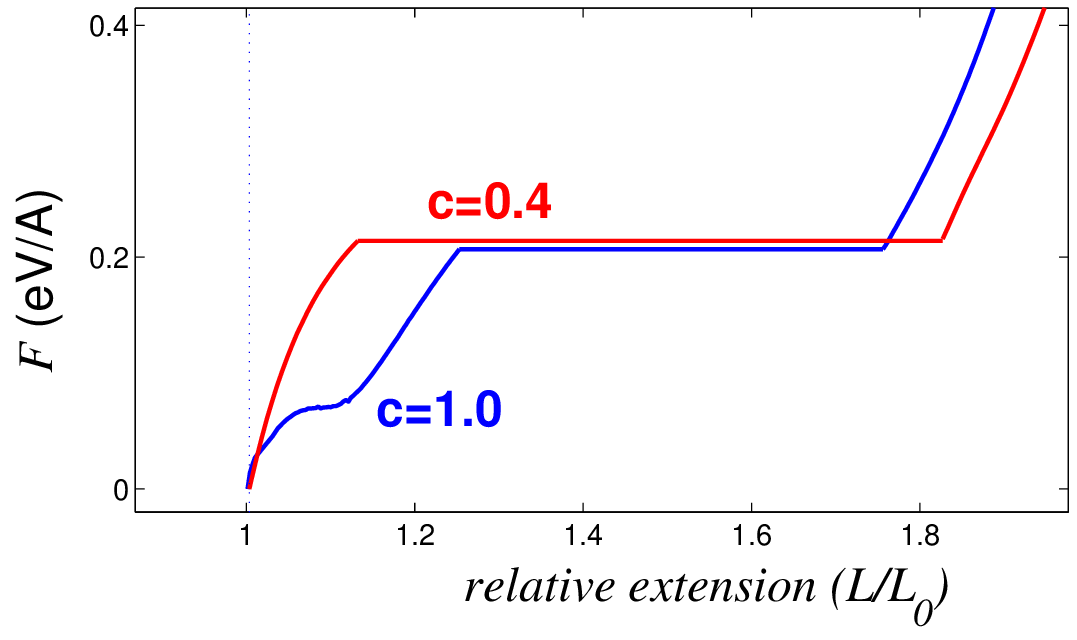}
\caption{\label{plateau_cmp}
Force-extension diagram for an ideal dsDNA helix. 
Blue line: artificially strong WC bonds ($c_0=1.0$).
Red line: regular strength WC bonds ($c_0=0.4$). Simulation at $T=300K$ 
}  
\end{figure}

\subsection{Insensitivity of the plateau transition to the WC bond strength}
Within the framework of 12CG coarse-grained model \cite{Savin2011} 
used in the paper, the relative strength of the WC hydrogen bonds is controlled 
by parameter $c_0$ in $E_{hb}^* = c_0 E_{hb}$, see section "Methods". 
We have used $c_0=0.4$ which gives an excellent 
agreement with the relevant 
single molecule stretching experiment \cite{vanMameren2009}, Fig.~\ref{figHB_AT}. 

Here we vary $c_0$ to test the effect that the hydrogen bond strength may have 
on the over-stretching plateau of double-stranded 
DNA. The main conclusion is that doubling the strength of the WC bonds -- 
to the point that they no longer break upon stretching at $300K$ -- 
has little effect on the existence of the over-stretching  plateau 
in the force-extension diagram. 
Specifically, we have performed a 700 ps long simulation 
of the same 500 b.p. poly(A)-poly(T) DNA fragment at  300K, but 
now with an unphysically large value of $c_0=1.0$ 
intended to keep the WC bonds from breaking. Now, even in the 
stretched state (see Fig.~\ref{fig_cmp_fHB}) the bonds do not break. However the plateau in the force-extension diagram still exists, 
(see Fig.~\ref{plateau_cmp}). Moreover, one can see from Fig.~\ref{plateau_cmp}
that the value of the tension at the plateau and its range 
differ only slightly from the $c=0.4$ case where the bonds do break as the 
chain is stretched, in perfect agreement with the experiment.  
Therefore, the existence and key characteristics 
of the plateau in the force-extension diagram 
for double-stranded DNA are rather insensitive to hydrogen bond strength. 
The comparison of the DNA stretching behavior in these two parameter 
regimes -- with regular and artificially strong WC bonds -- 
is another confirmation  
that  the DNA over-stretching plateau does not arise from WC bond breaking: 
the plateau exists even when the hydrogen bonds remain unbroken. 

\subsection{Computational resources}
Most of computationally intense calculations presented here such as 
minimization and molecular dynamics simulation of 
500 bp DNA were performed at Joint Supercomputer Center of Russian Academy of Science.


\end{document}